\documentclass[journal]{IEEEtran}

\usepackage{hyperref}
\hypersetup{hidelinks=true}

\usepackage{amssymb}
\usepackage{pifont}

\def\OJlogo{\vspace{-4pt}$<$Society logo(s) and publication title will appear here.$>$}
\def\seclogo{\vspace{10pt}$<$Society logo(s) and publication title will appear here.$>$}

\usepackage{graphicx}
\usepackage{float} 
\usepackage{textcomp} 
\usepackage[nospace,noadjust]{cite} 
\usepackage{amsmath,amssymb,amsfonts} 
\usepackage{amsthm} 
\usepackage{mathrsfs} 
\usepackage[mathscr]{euscript} 
\usepackage{epstopdf} 
\usepackage{xcolor}
\usepackage{pgfplots}
\pgfplotsset{compat=newest} 
\usepackage{tikz}
\usetikzlibrary{plotmarks}
\usepackage[normalem]{ulem} 
\usepackage{gensymb} 
\usepackage{balance}
\usetikzlibrary{intersections,backgrounds, patterns}
\usepackage{array}
\usepackage[nolist]{acronym} 
\usepgflibrary{arrows}
\usepackage{multirow} 

\usepackage{soul}

\usepackage{authblk}

\usepackage[cmintegrals]{newtxmath} 
\usepackage[font=footnotesize]{subcaption}

\usepackage[font=footnotesize]{caption}

\theoremstyle{remark}
\newtheorem*{remark}{Remark}

\usepackage{algorithm}
\usepackage{algorithmic}

\definecolor{mycolor1}{RGB}{19, 133, 189}
\definecolor{mycolor2}{RGB}{230, 112, 32}
\definecolor{mycolor3}{RGB}{130, 173, 98}
\definecolor{mycolor4}{rgb}{0.49412,0.18431,0.55686}%
\definecolor{myhist}{RGB}{73, 80, 87}
\definecolor{hgreen}{rgb}{0, 0.5, 0}

\pgfplotsset{every axis/.append style={
		scaled y ticks = false, 
		scaled x ticks = false,
		tick label style={/pgf/number format/fixed},
		label style={font=\footnotesize},
		tick label style={font=\footnotesize},
		tick scale binop=\times
	}
}

\begin{document}
\title{REM-U-net: Deep Learning Based Agile REM Prediction with Energy-Efficient Cell-Free Use Case}

\author[1]{Hazem Sallouha\thanks{Corresponding author: Hazem Sallouha (email: hazem.sallouha@kuleuven.be). This work was supported by the NSF under grant 2224322. The work of Hazem Sallouha was funded by the Research Foundation – Flanders (FWO), Postdoctoral Fellowship No. 12ZE222N, and he conducted this research while he was a Fulbright visiting scholar at the University of California, Los Angeles, USA.}}
\author[2]{Shamik Sarkar}
\author[2]{Enes Krijestorac}
\author[2]{Danijela Cabric}
\affil[1]{\normalsize{Electrical Engineering Department, KU Leuven, Leuven, Belgium}}
\affil[2]{\normalsize{Electrical and Computer Engineering Department, University of California, Los Angeles, USA}}

\maketitle

\begin{abstract}
Radio environment maps (REMs) hold a central role in optimizing wireless network deployment, enhancing network performance, and ensuring effective spectrum management. Conventional REM prediction methods are either excessively time-consuming, e.g., ray tracing, or inaccurate, e.g., statistical models, limiting their adoption in modern inherently dynamic wireless networks. Deep-learning-based REM prediction has recently attracted considerable attention as an appealing, accurate, and time-efficient alternative. However, existing works on REM prediction using deep learning are either confined to 2D maps or use a limited dataset. In this paper, we introduce a runtime-efficient REM prediction framework based on u-nets, trained on a large-scale 3D maps dataset. In addition, data preprocessing steps are investigated to further refine the REM prediction accuracy. The proposed u-net framework, along with preprocessing steps, are evaluated in the context of \textit{the 2023 IEEE ICASSP Signal Processing Grand Challenge, namely, the First Pathloss Radio Map Prediction Challenge}. The evaluation results demonstrate that the proposed method achieves an average normalized root-mean-square error (RMSE) of 0.045 with an average of 14 milliseconds (ms) runtime. Finally, we position our achieved REM prediction accuracy in the context of a relevant cell-free massive multiple-input multiple-output (CF-mMIMO) use case. We demonstrate that one can obviate consuming energy on large-scale fading measurements and rely on predicted REM instead to decide on which sleep access points (APs) to switch on in a CF-mMIMO network that adopts a minimum propagation loss AP switch ON/OFF strategy.
\end{abstract}

\begin{IEEEkeywords}
AP switch ON/OFF, cell-free, deep learning, large-scale fading, pathloss, radio environment map, received signal strength, spatial prediction, u-net
\end{IEEEkeywords}

\IEEEpeerreviewmaketitle

\begin{acronym}

\acro{AP}{access point}
\acro{ASO}{AP switch ON/OFF}
\acro{BW}{black and white}

\acro{LSF}{large-scale fading}
\acro{RF}{radio frequency}
\acro{REM}{radio environment map}
\acro{CPU}{central processing unit}
\acro{CNN}{convolutional neural network}

\acro{CF-mMIMO}{cell-free massive multiple-input multiple-output}
\acro{NLoS}{non-line-of-sight}
\acro{LoS}{line-of-sight}

\acro{MSE}{mean-square error}
\acro{MPL-ASO}{minimum propagation loss aware ASO}

\acro{SNR}{signal-to-noise ratio}
\acro{SE}{spectral efficiency}
\acro{GPU}{graphics processing unit}
\acro{KL}{Kullback–Leibler}
\acro{DNN}{deep neural network}
\acro{RSS}{received signal strength}
\acro{RMSE}{root-mean-square error}
\acro{UE}{user equipment}

\end{acronym}
\section{Introduction} \label{section:intro}
Predicting \acp{REM} that captures \ac{LSF} could facilitate deployment planning and operation optimization of wireless networks. For instance, REM prediction can play an essential role in spectrum sharing~\cite{achtzehn2012improving}, localization~\cite{sarkar2020llocus}, path planning for UAVs~\cite{krijestorac2021spatial}, finding coverage holes~\cite{azari2022evolution}, optimizing resource allocation in dense networks \cite{kulzer2021cdi}, etc. 

\textit{Why not ray tracing?} REM prediction using ray-tracing~\cite{rizk1997two} is well-known for its high accuracy, outperforming other conventional statistical model-based methods, such as the 3GPP spatial channel model~\cite{3gpp}, COST 231 model~\cite{damosso1999digital}, and WINNER II model~\cite{meinila2009winner}. However, the challenge with ray tracing is its prohibitively long computation time, limiting its adoption in modern networks, which are designed to be dynamic in terms of frequent resource allocation, reconfigurability, and mobility. This issue of computation time is especially critical for resource-constrained devices/nodes in distributed wireless networks.

\textit{Deep learning vs. ray tracing:} Recent research works have explored the use of supervised deep learning, specifically \acp{CNN} as a swift alternative for ray-tracing-based REM prediction~\cite{krijestorac2021spatial,levie2021radiounet,ratnam2020fadenet}. 
The basic idea is to use a \ac{DNN} as a function that approximates the input-output mapping of ray tracing methods in a faster manner. However, this faster approximation comes at the cost of reduced prediction accuracy. Accordingly, a significant focus of ongoing research in deep learning-based REM prediction is to improve the prediction accuracy while at the same time ensuring a bounded prediction runtime, e.g., in the scale of a few milliseconds. This research goal was also the foundation of the \textit{2023 IEEE ICASSP First Pathloss Radio Map Prediction Challenge}. At the same time, deep learning-based REM prediction has also been shown to be superior to traditional signal processing-based REM prediction~\cite{levie2021radiounet}.

\textit{Reactive vs. proactive deep learning REM prediction:} Deep learning-based REM prediction, as well as REM prediction in general, can be broadly categorized as reactive and proactive. Reactive REM prediction relies on a small set of RSS measurements from an active transmitter whose radio environment is to be predicted~\cite{krijestorac2021spatial}. In contrast, proactive REM prediction is capable of making predictions for transmitters for which no measurements are available~\cite{levie2021radiounet,sarkar2023prospire}. Both of these approaches for REM prediction have their own challenges and benefits. For example, when planning for deployment of base stations (BS) in a geographical area, reactive REM prediction is less practical as there are no active transmitters from which sparse RSS measurements can be collected\footnote{One way to deal with this limitation is wardriving with mobile BSs and mobile receivers; however, that is practically inconvenient.}. In contrast, proactive REM prediction is much more convenient for this problem as no RSS measurement is required. However, the lack of RSS measurements is also a significant challenge with proactive REM prediction. Specifically, what should be the basis for REM predictions? Most recent works in deep learning-based proactive REM prediction have relied on training data collected from various geographical areas (not including the target areas where predictions will be made in the online/testing phase) via ray tracing. It is important to note that using ray tracing for collecting the training data is not an issue because the time needed to collect/generate training data does not affect the real-time operation in the online phase. 

\textit{Goal of our work:} Given the broader appeal of proactive REM prediction, in this paper, we consider the problem of predicting signal strength across an area due to a transmitter at a given location within the area. One of our primary motivations for investigating this problem was to address one of the \textit{2023 IEEE ICASSP Signal Processing Grand Challenges}, namely, the \textit{First Pathloss Radio Map Prediction Challenge}~\cite{REMPredictionChallenge}. Unlike existing works in the literature, which are based on either relatively limited dataset~\cite{ratnam2020fadenet}, simulated maps~\cite{krijestorac2021spatial}, or 2D maps~\cite{levie2021radiounet}, the dataset considered in this challenge is a large-scale 3D dataset. This 3D dataset consists of over $701$ geographical areas (henceforth city maps) with varying numbers of transmitters and building heights for each of the city maps. More details about this dataset can be found in~\cite{radioDataset} and are also described later in Section~\ref{section:system_model}.

\textit{Our approach:} To address the above problem, in this work, we develop an approach that relies on the u-net~\cite{ronneberger2015u} neural network architecture. While u-net has been used in similar problems, we develop new strategies for adopting u-net for our problem with a 3D dataset. Additionally, we share several insightful findings that we discovered while participating in the radio map prediction challenge. First, we showed that using \ac{LoS} information as an input to the neural network leads to better prediction accuracy. However, precomputing the LoS information incurs additional computation time, which hinders the primary goal of deep learning-based fast REM prediction. Hence, we developed three different methods for computing LoS information in our approach. These methods differ in terms of the quality of LoS information and computation time. 
Second, during our experiments, we learned that the density of buildings in the target area impacts the prediction accuracy. Hence, it is useful to train two different deep learning models: one based on city maps with low density of buildings and another based on city maps with high density of buildings. During the online phase, we can choose one of these two models based on the density of buildings in the target area. The impact of buildings' density has a higher impact, especially when the amount of training data is limited. 
Third, we identified that instead of training the neural network to predict the average signal strength, it might be helpful to train the neural network (actually two neural networks, as explained later) to predict the probability distribution of the signal strength. We showed that in certain scenarios, this approach can lead to better prediction accuracy. Additionally, as discussed in~\cite{krijestorac2021spatial}, predicting the distribution of signal strength can have additional benefits in specific applications.

Based on our learned lessons and insights, we used one particular combination of our developed strategies as our solution to the \textit{ICASSP First Pathloss Radio Map Prediction Challenge}. Specifically, we used the buildings and transmitters location and height information, along with \ac{LoS} maps, stacked together as input to two u-nets to predict the probability distribution (mean and variance) of the signal strength. These two u-nets follow our third insight described above, in which we model the signal strength as a Gaussian random variable and train the two u-nets to predict the probability distribution (mean and variance) of the signal strength. The evaluation results, with unseen city maps, show that our proposed method provides an average normalized \ac{RMSE} of 0.045 with an average runtime of 14 milliseconds. Our approach and results for the \textit{ICASSP First Pathloss Radio Map Prediction Challenge} are briefly summarized in~\cite{krijestorac2023agile}. However, in this paper, we share additional results that are not presented in~\cite{krijestorac2023agile}. 

In order to position our achieved \ac{REM} prediction accuracy in the context of a relevant application, we consider the minimum propagation loss \ac{ASO} strategy \cite{femenias2020access} proposed to address the excessive power consumption concerns in \ac{CF-mMIMO} networks. The essence of the \ac{MPL-ASO} strategy is to activate only a subset of \acp{AP} that is sufficient to meet \acp{UE} \ac{SE} based on the pathloss gain between \acp{AP} and \acp{UE} \cite{femenias2020access} and set the rest of \acp{AP} in sleep mode. In order to enable \ac{ASO} strategies, existing works in the literature assume that all \acp{AP} are frequently turned on to collect channel measurement, including \ac{LSF} \cite{femenias2020access,garcia2020energy,van2020joint}. Alternatively, our \ac{CF-mMIMO} use case demonstrates that predicted \acp{REM} of the off \acp{AP} can be used in the \ac{AP} selection problem of \ac{MPL-ASO}, improving \ac{CF-mMIMO} networks energy-efficiency by eliminating the need to frequently turn \acp{AP} on. Our results show that by exploiting predicted \ac{REM}, we attain an \ac{AP} selection error of around 5\% in case a \ac{UE}'s \ac{SE} needs three extra \acp{AP}.

\subsection{Contributions} In summary, our contributions in this paper are the following:
\begin{itemize}
    \item We present three different LoS calculation methods that rely only on a given transmitter location and the corresponding 3D city map, offering a tradeoff between accuracy and calculation time. In particular, these methods are per-pixel calculation, accelerated batch calculation, and neural-network-based calculation, all detailed in Section~\ref{section:methods}. Unlike existing works in the literature, which only exploit binary LoS maps, our LoS maps present a fractional value for \ac{NLoS} pixels, depending on the number of encountered buildings. This domain-knowledge-based information assists the neural network to learn in a swift and accurate manner. We quantify the performance gain of these three LoS calculation methods individually when utilized as a preprocessing step for REM prediction.
    \item We propose a u-net-based \ac{CNN} to predict REM using transmitter location and city map information. We explore, in addition to LoS, also building density split and data augmentation as preprocessing steps. Furthermore, we present several insightful findings that we discovered while participating in the radio map prediction challenge. In particular, these findings are 1) the performance gain when using LoS maps as an input to the neural network, 2) the positive impact of training two models based on the building density when the training dataset is limited, and 3) the performance gain obtained when training a u-net to predict the probability distribution instead of the average signal strength. We quantify these performance gains and impact using the \textit{RadioMap3DSeer} dataset. 
    \item We introduce a novel energy-efficient \ac{CF-mMIMO} use case with REM-prediction-based \ac{MPL-ASO}. We show that by relying on REMs predicted using our proposed u-net, along with data augmentation and LoS preprocessing steps, we eliminate the need to frequently turn sleep APs on to do channel measurements needed for \ac{MPL-ASO}. We evaluate the performance of our proposed REM-prediction-based \ac{MPL-ASO}, showing that it achieves an \ac{AP} selection error of approximately 5\% when using the predicted REM compared to the true one.
\end{itemize}

\subsection{Organization} The rest of the paper is organized as follows. In Section~\ref{section:related_work}, we discuss the relevant related works. Next, in the Section~\ref{section:unet_primer}, we present a primer on u-net. Section~\ref{section:system_model} presents our system model and describes the dataset used in this paper. We present our proposed methods in Section~\ref{section:methods}, and the corresponding evaluation results in Section~\ref{sec:REMresults}. Next, we present a use case of REM prediction, specifically, \ac{AP} ON/OFF switching in \ac{CF-mMIMO}, in Section~\ref{sec:CF}. Finally, Section~\ref{sec:conclusion} provides the conclusions.

\section{Related Work} \label{section:related_work}
When sparse RSS measurements are available from the targeted active transmitter, i.e., for the case of reactive REM prediction, several methods have been investigated. The simplest method is to perform the predictions as a weighted average of the available measurements. For instance, in inverse distance weighting (IDW) methods, the weighting is done using a heuristic approach based on the inverse of the distance between the target location and the measurement location~\cite{sarkar2020llocus}. Another weighted-average-based example is the Kriging interpolation method, which uses a weighted average of the measured RSS, and obtains weights based on an optimization approach~\cite{maeng2022out,chakraborty2017specsense}.
Alternatively, several works investigated the usage of deep learning for reactive REM prediction based on sparse measurements~\cite{teganya2021deep,li2021model,krijestorac2021spatial}. These works transformed the available information, e.g., sparse RSS measurements and their associated locations, transmitter locations, and environment map, into a set of stacked images (i.e., tensor) and fed it to a \ac{DNN} to predict the REM. The deep neural networks used have an encoder-decoder structure, e.g., u-net~\cite{krijestorac2021spatial}, autoencoders~\cite{teganya2021deep}, ResNets~\cite{li2021model}, so that the problem can be formulated as an image-to-image translation. In general, most of these works show the benefits of using deep learning over signal processing and heuristics methods, either in terms of REM prediction accuracy or prediction time. 

On the other hand, when sparse measurements are unavailable for the targeted transmitter, i.e., for the case of proactive REM prediction, the most accurate REM prediction method is ray tracing~\cite{rizk1997two}. However, as discussed in Section~\ref{section:intro}, the computation time is the major drawback of ray tracing methods. The simplest way to perform the prediction is to use the free space radio wave propagation pathloss model that maps RSS to the distance from the transmitter~\cite{rappaport1996wireless}. However, this method does not take into account the shadow fading. Hence, various statistical modeling-based approaches have been developed, e.g., the log-normal shadowing model \cite{goldsmith2005wireless}. However, such radial-symmetric statistical methods, fail to capture the variation from one environment to another, e..g, location and height of buildings/obstacles.

An alternative promising way of performing proactive REM prediction is to collect training data via ray tracing from different cities/areas to learn a deep learning model that can perform predictions on unseen areas or transmitters~\cite{levie2021radiounet,ratnam2020fadenet}. In \cite{ratnam2020fadenet}, the authors proposed a variant of the u-net architecture to predict the \ac{LSF} maps of mmWave base stations. The authors used ray tracing to generate the training and ground-truth LSF maps from only three cities, considering terrain, transmitter location, buildings, foliage, as well as \ac{LoS} information as input. A u-net-based method is introduced in \cite{levie2021radiounet} to predict the \ac{REM} of unseen transmitters in a 2D scenario, with training data constructed from buildings maps and transmitters locations. In this work, the authors used maps of buildings layout from 6 cities, following the dataset presented in \cite{radioDataset}. However, buildings, transmitters, and receivers are each set to constant heights across all training and testing data. The promising potential of u-net-based REM prediction and the substantial importance of filling the literature gaps regarding the lack of height information and the dataset size limitations inspire our work in this article.

\section{Primer on U-net} \label{section:unet_primer}
In this section, we briefly present a primer on u-net which is the core of our proposed approach in this paper.

\acp{CNN} are arguably the most popular deep learning network architecture.
\acp{CNN} are widely adopted in image processing tasks, demonstrating exceptional capabilities in image classification and object detection problems \cite{li2021survey}. However, for pixel-wise image segmentation and regression, a different neural network architecture is required compared to those commonly used for image classification and object detection.
For such problems, a special \ac{CNN} architecture, known as u-net, has emerged as a promising solution~\cite{ronneberger2015u}.

The u-net architecture, introduced in \cite{ronneberger2015u}, consists of an encoder contracting path, a decoder expansive path, and skip connections. In the \textit{encoder path}, convolution and pooling operations are used to progressively reduce the spatial dimensions of the input while increasing the number of feature channels. This process helps in extracting hierarchical representations of the input image. The \textit{decoder path} of the u-net is responsible for upsampling the feature maps back to the original input size. In this path, transpose convolutions (also known as deconvolutions or upsampling) are used to progressively increase the spatial dimensions and reduce the number of feature channels. Since the upsampling in the decoder path lacks high-resolution information, \textit{skip connections} are employed between the encoder and the decoder paths, copying and concatenating the feature maps in the encoder layers to the corresponding feature maps of the decoder layers. The skip connections help preserve spatial information and promote better segmentation. The resulting architecture is U-shaped, hence the name u-net.

\section{System Model and Dataset Structure} \label{section:system_model}
In this section, we introduce the system model, highlighting both the environment description and network assumptions. Subsequently, we present the details of the dataset used in this work.
\begin{figure}[t]
	\centering
	\includegraphics[width=0.45\textwidth]{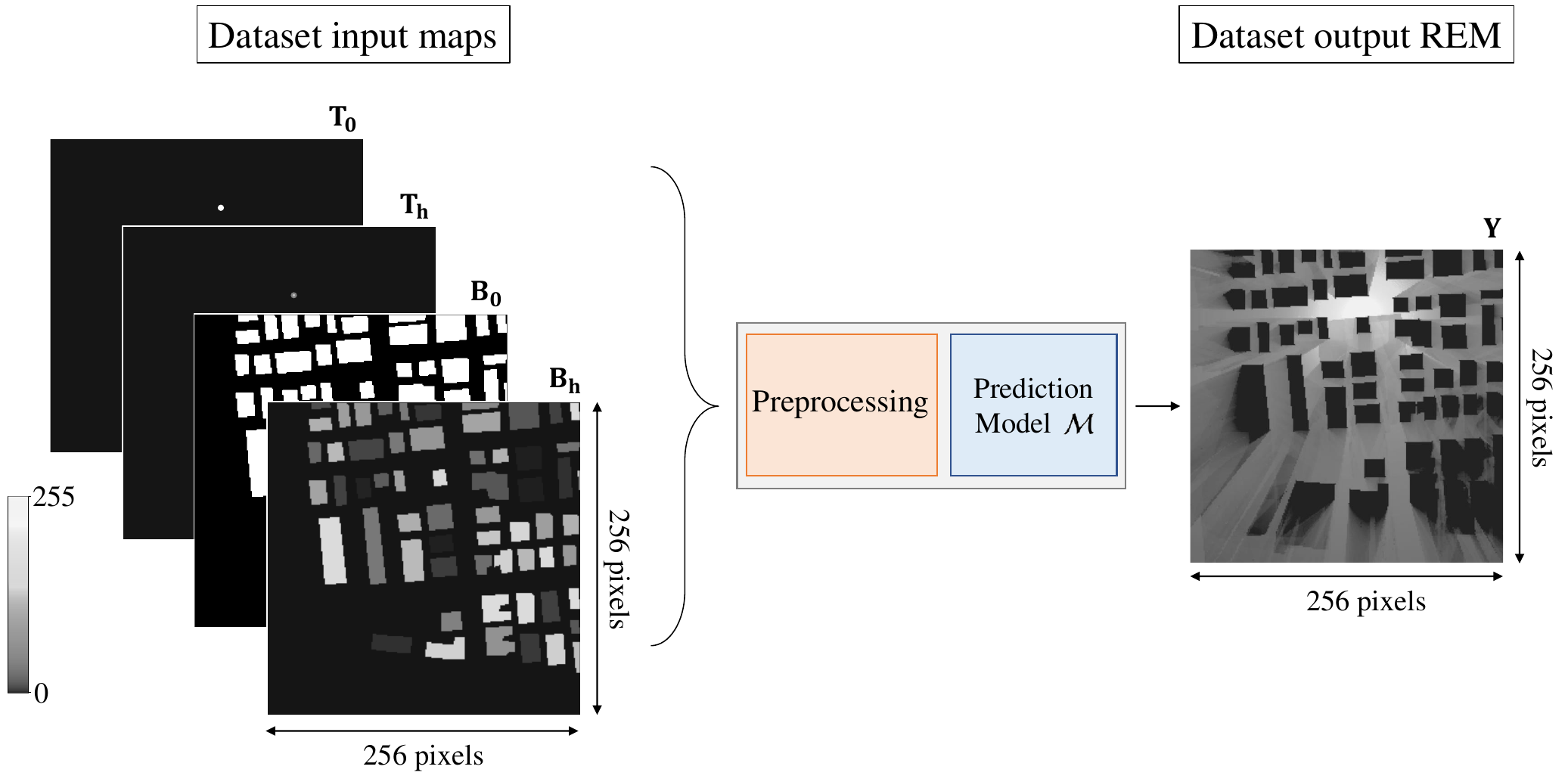}
	\caption{A visualization of the considered REM prediction system, showing a sample input and the corresponding REM output. Our objective is to define data preprocessing steps and design a prediction model $\mathcal{M}$.}
	\label{REM_model}
\end{figure}

\subsection{System Model} 
The propagation environment considered in this work is an outdoor urban environment, with relatively narrow inter-building space and limited building heights ranging between 2 to 6 stories/floors. Transmitters, which are also known as base stations or \acp{AP}, are placed on top of the buildings, whereas \ac{UE}s are assumed to be on the ground at a constant 1.5 m height. Single omnidirectional antennas are assumed to be used by both transmitters and receivers, working at the center carrier frequency of 3.5 GHz with a system bandwidth of 20 MHz. 

\textit{Problem Statement:} Our main objective is to design a model $\mathcal{M}$, along with any companion preprocessing steps needed, taking 3D map information and transmitter locations as input and predicting the corresponding \ac{REM} as an output, as depicted in Figure \ref{REM_model}. In addition to the high prediction accuracy, we are also aiming at minimizing the prediction run-time of the model $\mathcal{M}$ to be in the scale of milliseconds.   

\subsection{Dataset Structure and Settings}
In this work, we use the \textit{RadioMap3DSeer} dataset \cite{radioDataset}, which consists of 701 city maps with buildings layouts fetched from \textit{OpenStreetMap} \cite{OpenStreetMap} of 6 urban European-style cities, e.g., London, and Berlin. Intelligent ray tracing (IRT) \cite{hoppe1999fast} is used to simulate the \acp{REM}, using \textit{WinProp} \cite{hoppe2017wave} software, assuming IRT with a maximum of two interactions (diffractions and/or reflections). Simulations were conducted for 80 different transmitter locations per map, resulting in a total of 56080 \acp{REM}. Each map is 256\,$\times$\,256 m$^2$, stored as an image with a resolution of 256\,$\times$\,256 pixels, implying that each pixel represents 1\,$\times$\,1 m$^2$. In particular, as shown in Figure \ref{REM_model}, dataset maps include:

\begin{table}[t]
	\caption{\textit{RadioMap3DSeer} dataset Parameters.}
	\centering
	\begin{tabular}{ | l || c |}
		\hline
		\textbf{Parameter} & Value\\
		\hline
		\hline
		Single map size & 256\,$\times$\,256 m$^2$\\
		\hline
		Simulation environment & Urban\\
		\hline
		Carrier frequency &  3.5 GHz \\
		\hline
		System bandwidth &  20 MHz \\
		\hline
		Transmit power & 23 dBm \\ 
		\hline
		Antenna radiation pattern & Omnidirectional \\
		\hline
		Maximum path gain & $- 75 $ dB \\
		\hline
		Minimum truncated path gain & $- 111 $ dB \\
		\hline
		Buildings height range & 6.6 - 19.8 m \\
		\hline
		Transmitter height & 3 m above rooftop \\
		\hline
	\end{tabular}
	\label{DS_para}
\end{table}

\begin{itemize}
	\item \textit{Buildings layout}: Each of the 701 buildings layout maps is provided as a binary image as well as an image with quantized building heights. In the binary image, $\textbf{B}_0$, pixels of buildings are set to ones and the rest to zeros, whereas in the quantized buildings height map, $\textbf{B}_h$, pixels of buildings show the quantized height of buildings. Each building in a city is assigned a height ranging from 2 to 6 stories, with a constant story height of 3.3 m, resulting in buildings' heights ranging from 6.6 m to a maximum of 19.8 m. The buildings' heights were stored in $\textbf{B}_h$ as uniformly quantized values between $[1, 255]$.    

	\item \textit{Transmitters locations}: Transmitters are assumed to be placed on buildings' rooftops, considering only buildings with a minimum height of 16.5 m located within the 150×150 center area to accommodate a transmitter. The transmitters are placed at the building edge with a height of 3 m from the corresponding building rooftop. Concerning each transmitter location, the dataset contains two 256\,$\times$\,256 pixels images. The first image is a binary image denoted by $\textbf{T}_0$, in which only the pixel containing the transmitter is set to one, and in the other image denoted by $\textbf{T}_h$, only the pixel containing the transmitter presents the value of the corresponding building height.
	
	\item \textit{Radio environment maps}: \acp{REM} are simulated using \textit{WinProp} \cite{hoppe2017wave} software and path gain values of each 1\,$\times$\,1 m$^2$ are stored in dB scale, considering a constant transmit power of 23 dBm. The maximum reported path gain is $-75$ dB and all path gain values below the analytical threshold of $-111$ dB are truncated, providing a path gain range of 36 dB $(-75 - (-111))$. Each \ac{REM}, denoted by $\textbf{Y}$, is scaled to gray levels between 0 and 255, enabling the authors \cite{radioDataset} to save \acp{REM} as images.
\end{itemize}
Table \ref{DS_para} summarizes the main settings and parameters of the \textit{RadioMap3DSeer} dataset \cite{radioDataset} considered in this work.

\section{Agile Radio Map Prediction} \label{section:methods}
In this section, we present our proposed approach. First, we present the architecture of u-net that we use in all of our methods unless otherwise stated. Then, we describe the different possible preprocessing steps that we investigate in this paper. Finally, we present an alternative neural network architecture that we used as part of the ICASSP challenge.

\subsection{U-net architecture} \label{subsection:unet_arch}
\begin{figure}[t]
	\centering
	\includegraphics[width=0.48\textwidth]{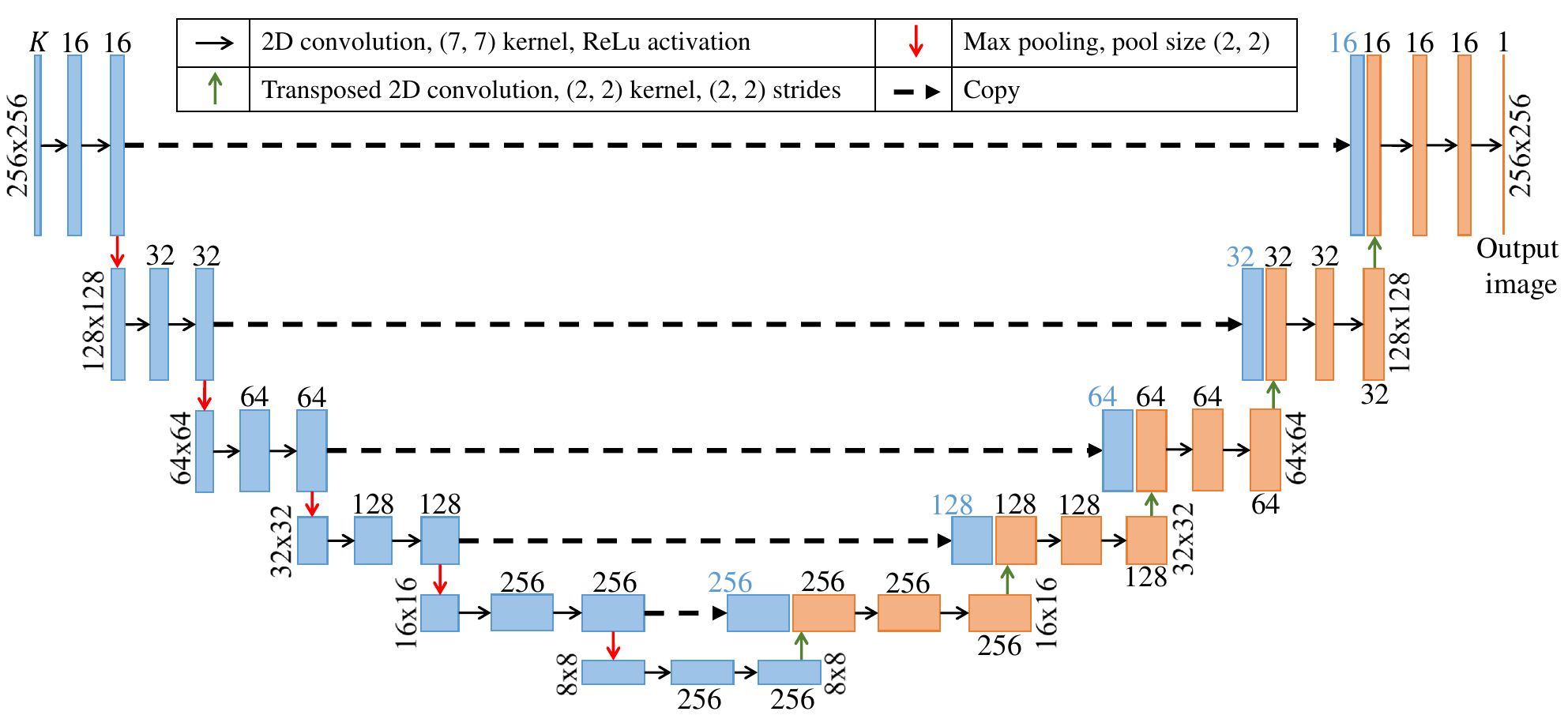}
	\caption{A detailed architecture of the U-net used in this work.}
	\label{fig:unet_arch}
\end{figure}

As discussed in Section~\ref{section:unet_primer}, u-net architecture can achieve superior segmentation performance. In the context of our problem, the REM prediction is similar to segmentation, with the only difference being that for each of the output pixels, we need to perform regression instead of classification. This idea has been explored in some prior works on REM prediction and serves as our motivation for using u-net. The u-net architecture uses an image-to-image translation approach. Hence, the available information for making the REM predictions must be transformed into images. Conveniently, the available information is encoded as images in the dataset described in Section~\ref{section:system_model} (refer to $\textbf{B}_0$, $\textbf{B}_h$, $\textbf{T}_0$, and $\textbf{T}_h$). We stack several images as a 3D tensor and feed it to the u-net. Accordingly, the operation of the u-net can be formally described as the following mapping $\mathcal{M}: \mathcal{R}^{256 \times 256 \times K} \rightarrow \mathcal{R}^{256 \times 256}$. The value of $K$ depends on the number of images we use as input. The architecture of u-net that we primarily rely on is shown in Figure~\ref{fig:unet_arch}. We train $\mathcal{M}$ by minimizing the mean squared error between the true REM $\textbf{Y}$ and the estimated REM $\hat{\textbf{Y}}$.

\subsection{Using LoS information as additional input to u-net} \label{sec:los_compute_methods}
\begin{figure}[t]
	\centering
	\includegraphics[width=0.45\textwidth]{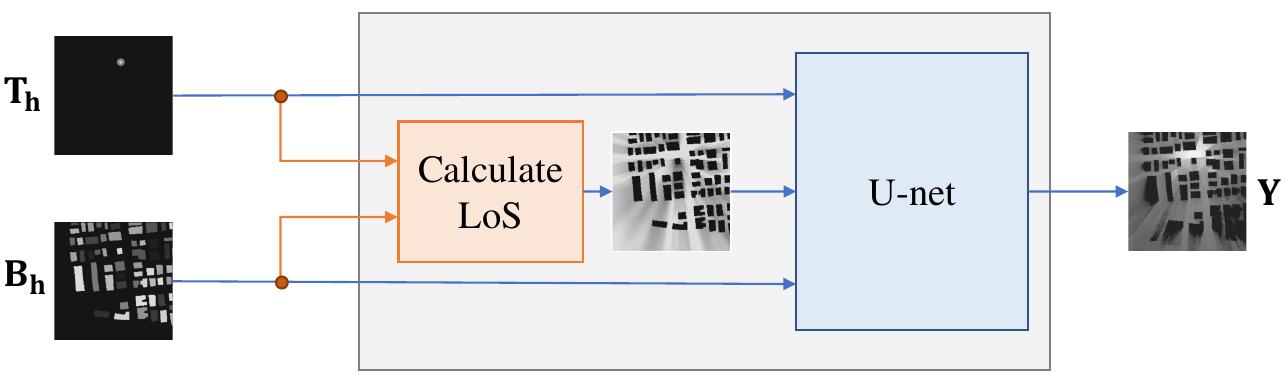}
	\caption{A visualization of the proposed U-net model assisted by providing $\textbf{L}_f$ as an additional input, denoted by $\mathcal{M}(\text{noDAug}, \text{LoS}_f, \text{U-net}, \text{MSE})$.}
	\label{LosModel}
\end{figure}

In addition to using the available information from the dataset as input to our neural network, we use an additional input, which we call the LoS map, $\textbf{L}_f$. Specifically, in this section, we use $K=3$ and the three input images to the u-net are $\textbf{T}_h$, $\textbf{B}_h$, and $\textbf{L}_f$. As discussed next, we compute $\textbf{L}_f$ using $\textbf{T}_h$ and $\textbf{B}_h$. Since both $\textbf{T}_h$ and $\textbf{B}_h$ are used as input to the neural network, the neural network should be able to learn and perform equally well with and without using $\textbf{L}_f$ as an additional input. However, the learning task of the neural network without $\textbf{L}_f$ would be more complex. Hence, based on our domain knowledge of radio wave propagation, we assist the neural network in learning quickly by providing $\textbf{L}_f$ as an additional input, as illustrated in Figure \ref{LosModel}. We show later in Section \ref{sec:REMresults} that, indeed, precomputing the LoS map and using it as an input to the neural network improves the REM prediction accuracy. In the following, we describe three different methods for computing $\textbf{L}_f$.

\subsubsection{Per-pixel LoS calculation (PxLoS)} \label{subsubsection:per_pixel_los}
For a given transmitter location, $(x_t, y_t, z_t)$, we compute the LoS information for each of the pixels in the area where the transmitter is located. First, for a particular pixel, say $(x_r, y_r)$, we form a straight line, $l_{tr}$, in the 3D space between $(x_t, y_t, z_t)$ and $(x_r, y_r, z_r)$. Note that $(x_r, y_r)$ denote the pixel center in the XY plane and $z_r = 1.5$ m as receivers are assumed to be always 1.5 m above the ground level as discussed in Section~\ref{section:system_model}. The straight line $l_{tr}$ is defined as an ordered sequence of $N_l$ coordinates $\big \langle(x_l, y_l, z_l)\big \rangle; l = 1, 2, ..., N_l$, where $N_l$ is the number of pixels that $l_{tr}$ passes through. The set of pixels, $\mathcal{P} = \{ (x_l, y_l)\}; l = 1, 2, ..., N_l$, associated with $l_{tr}$ are found using the Bresenham’s line algorithm~\cite{bresenham}. In this method, we move from $x_r$ to $x_t$ one pixel at a time and find the pixels along $y$ dimension based on the slope of the line such that the sequence of pixels in $\mathcal{P}$ closely approximates a straight line between $(x_r, y_r)$ and $(x_t, y_t)$. For computing the values of $z_l$, we compute the slope along $z$ dimension as $\frac{z_t - z_r}{x_t - x_r}$ and use $z_l = z_r + (x_l - x_r) \times \frac{z_t - z_r}{x_t - x_r}$. Then, for each of the pixels in $\mathcal{P}$, we check the building height for that pixel from $\textbf{B}_h$. For a particular pixel in $\mathcal{P}$, if the building height is less than $z_l$ (attribute of $l_{tr}$), we say that the pixel is in LoS. Based on this checking, we form another set, $\mathcal{L}$, that has the same size as $\mathcal{P}$ and has a one-to-one association with the pixels in $\mathcal{P}$. The elements of $\mathcal{L}$ are either 0 or 1, depending on whether the corresponding pixel is in LoS or not. 
Finally, for $\textbf{L}_f$, we set the value of pixel $(x_r, y_r)$ to be $\Big(1 - \frac{\sum_{i \in \mathcal{L}} \mathbf{1}_i}{|\mathcal{L}|}\Big)$ where $\mathbf{1}_i$ is an indicator function. $\mathbf{1}_i = 1$ if the $i^{th}$ element of $\mathcal{L}$ is 1, and  0 otherwise.

The primary disadvantage of the PxLoS method is that it is very slow as it performs the LoS check pixel by pixel for each transmitter. Specifically, if the target area has $P \times P$ pixels, then the complexity of generating  or  is $\mathcal{O}(P^3)$. 

\subsubsection{Accelerated batch LoS calculation (AbLoS)} \label{subsubsection:ablos}
The primary reason for the slowness of PxLoS is that the elements of $\textbf{L}_f$ are computed sequentially. To avoid that, in AbLoS, we simultaneously compute all the elements of $\textbf{L}_f$. Computing $\textbf{L}_f$ in one shot is especially attractive because we can use libraries that can vectorize the operations in AbLoS and accelerate the computation of $\textbf{L}_f$ in GPU. The details of this approach are described next.

This approach is also based on Bresenham’s algorithm. The basic idea of Bresenham’s algorithm is to find the set of \textit{n}-D pixels that closely approximate a \textit{n}-D line between two points. In our problem, \textit{n} is 3. For a given transmitter location, $(x_t, y_t, z_t)$, first, let us denote the set/batch of target 3D pixels for which LoS information must be computed as $\mathcal{B}$. Each of the elements of $\mathcal{B}$ is defined by $(x_r, y_r, z_r)$, where $(x_r, y_r)$ denote the pixel center in the XY plane and $z_r$ is always 0. Using $z_r=0$ instead of 1.5 m is an approximation. Next, based on Bresenham’s algorithm, we move from $(x_r, y_r, z_r)$ to $(x_t, y_t, z_t)$ one pixel at a time along the driving dimension. Here, driving dimension is the dimension among $x$, $y$, and $z$ that has maximum absolute difference between $(x_r, y_r, z_r)$ and $(x_t, y_t, z_t)$, i.e., $\max \{ |x_t - x_r|, |y_t - y_r|, |z_t - z_r| \}$. For example, if the driving dimension is $x$, then we traverse $x_r, x_r +1, ..., x_t$. While we move along the driving dimension, we also keep updating the other two dimensions, using pixel-wise increments based on the slope of the line between $(x_r, y_r, z_r)$ and $(x_t, y_t, z_t)$. For example, if the driving dimension is $x$, we define the slope along $y$ dimension as $\frac{y_t - y_r}{|x_t - x_r|}$ and the slope along $z$ dimension as $\frac{z_t - z_r}{|x_t - x_r|}$. Although the slope can be fractional, the increment along non-driving dimensions is always in integers. This is done by rounding off fractional values to the nearest integers. This procedure gives us a set of 3D pixels that closely approximate a line between $(x_r, y_r, z_r)$ and $(x_t, y_t, z_t)$. Importantly, in AbLoS, we assume that the distance to $(x_t, y_t, z_t)$ along the driving dimension from all the pixels in $\mathcal{B}$ are the same. Specifically, we assume this distance, say $D$, to be the maximum absolute difference between any pair $(x_r, y_r, z_r)$ and $(x_t, y_t, z_t)$, along any dimension ($x, y, z$). I.e., $D = \max_{B} \max \{ |x_t - x_r|, |y_t - y_r|, |z_t - z_r| \}$. This allows us to find the approximate lines, as explained before, between all the elements of $\mathcal{B}$ and $(x_t, y_t, z_t)$ simultaneously. Consequently, this vectorized operation can be significantly accelerated on a GPU using the software library CuPy \cite{cupy_learningsys2017}. Next, for each of the computed lines, we check the number of constituent 3D pixels that intersect with the buildings. This can be done by converting $\textbf{B}_h$ into a set of 3D pixels filled with 1s and 0s depending on whether a pixel is under a building or not and finding its intersection with the computed lines. Let us denote $p_r$ as the number of such intersecting 3D pixels for the line between $(x_r, y_r, z_r)$ and $(x_t, y_t, z_t)$.
This operation can also be accelerated as all the lines comprise the same number of 3D pixels. Finally, for $\textbf{L}_f$, we set the value of pixel $(x_r, y_r)$ to be $\Big(1 - \frac{p_r}{D}\Big)$. 

AbLoS uses an approximation that all the lines are made up of the same number of 3D pixels. However, it is much faster than PxLoS. In the above description, we have mentioned a couple of crucial operations where GPU-based acceleration is useful. Additionally, in our implementation, we leverage CuPy to accelerate other operations whenever possible. 

\subsubsection{Predicting LoS map via neural network (NNLoS)} \label{subsubsection:nnlos}
In this method, which we call NNLoS, we predict the LoS map instead of computing it. Specifically, we train a neural network that can approximate the LoS maps generated by the PxLoS method. 

During the training phase, for each of the training examples in the dataset, first, we compute $\textbf{L}_f$ using the PxLoS method. These LoS maps become the labels for training the neural network in NNLoS. Recall from Section~\ref{subsubsection:per_pixel_los} that for computing $\textbf{L}_f$ we need $\textbf{T}_h$ and $\textbf{B}_h$. Hence, we use $\textbf{T}_h$ and $\textbf{B}_h$ stacked as a 3D tensor as input to the neural network and train it to predict $\textbf{L}_f$. Let us denote this LoS predictor neural network as a function, $\mathcal{F}_{LoS}$, that performs the following mapping $\mathcal{F}_{LoS}: (\textbf{T}_h | \textbf{B}_h ) \rightarrow \textbf{L}_f$, where $|$ denotes the depth-wise stacking. Since $\mathcal{F}_{LoS}$ performs an image-to-image translation, we use a u-net to represent this function too. For the $\mathcal{F}_{LoS}$ u-net, we use the same neural network architecture as that for signal strength prediction (refer to Figure~\ref{fig:unet_arch}). We train $\mathcal{F}_{LoS}$ by minimizing the mean squared error between $\textbf{L}_f$ and its estimation, $\hat{\textbf{L}}_f$ using the Adam optimizer with learning rate of $10^{-4}$.

During prediction, we simply use the trained neural network, i.e., $\mathcal{F}_{LoS}$ to predict the LoS map $\hat{\textbf{L}}_f$. The computation time of $\hat{\textbf{L}}_f$ is, in general, much lower than computing the actual LoS map, $\textbf{L}_f$, via PxLoS. However, since $\hat{\textbf{L}}_f$ is an estimation of $\textbf{L}_f$, it negatively affects the REM prediction, as shown later in Section \ref{sec:REMresults}.
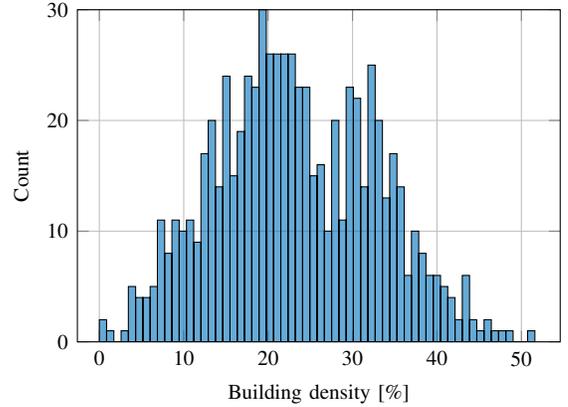
\begin{figure}[t]
	\centering
%
%
\definecolor{mycolor1}{rgb}{0.00000,0.44700,0.74100}%
\begin{tikzpicture}

\begin{axis}[%
	legend style={font=\fontsize{6}{5}\selectfont},
	width=8cm,
	height=6cm,
	xmin=-2.6,
	xmax=54.6,
	xlabel={Building density [\%]},
	ymin=0,
	ymax=30,
	ylabel={Count},
	axis background/.style={fill=white},
	xmajorgrids,
	ymajorgrids
]
\addplot[ybar interval, fill=mycolor1, fill opacity=0.6, draw=black, area legend] table[row sep=crcr] {%
x	y\\
0	2\\
0.86	1\\
1.72	0\\
2.58	1\\
3.44	5\\
4.3	4\\
5.16	4\\
6.02	5\\
6.88	11\\
7.74	8\\
8.6	11\\
9.46	10\\
10.32	11\\
11.18	9\\
12.04	17\\
12.9	20\\
13.76	14\\
14.62	24\\
15.48	15\\
16.34	19\\
17.2	24\\
18.06	23\\
18.92	30\\
19.78	26\\
20.64	26\\
21.5	26\\
22.36	26\\
23.22	23\\
24.08	23\\
24.94	15\\
25.8	16\\
26.66	10\\
27.52	20\\
28.38	11\\
29.24	23\\
30.1	22\\
30.96	14\\
31.82	25\\
32.68	20\\
33.54	13\\
34.4	17\\
35.26	14\\
36.12	6\\
36.98	10\\
37.84	8\\
38.7	6\\
39.56	6\\
40.42	5\\
41.28	4\\
42.14	2\\
43	6\\
43.86	2\\
44.72	1\\
45.58	2\\
46.44	1\\
47.3	1\\
48.16	1\\
49.02	0\\
49.88	0\\
50.74	1\\
51.6	1\\
};
\end{axis}
\end{tikzpicture}%
	\caption{A histogram of the building density (no. of buildings pixels divided by the total number of pixels).}
	\label{buildDens}
\end{figure}

\subsection{Environment building density} 
Radio channels are commonly modeled based on the characteristics of the propagation environment by either using different models for different environments, or a single model with different environment-dependent parameter values. For instance, in the log-normal pathloss model, the pathloss exponent has different values for environments such as urban, suburban, indoor, etc.~\cite{goldsmith2005wireless}. A key element that characterizes outdoor propagation environments is the building density, which influences both small-scale multipath fading as well as large-scale shadow fading. Intuitively, the buildings' density noticeably varies when comparing dense-urban, urban, or suburban environments.
Inspired by the influence buildings' density has on radio wave propagation, in this section, we explore the building density in the 701 maps provided in the \textit{RadioMap3DSeer} dataset. We define building density in a given city map as the number of buildings pixels divided by the total number of pixels. Figure \ref{buildDens} presents the histogram of the buildings' density per map. As shown in the figure, a bimodal distribution with two peaks can be spotted around 20\% and around 31\%. These two peaks imply that there are two different groups of building density, which can be translated to two different propagation environments, e.g., dense-urban and urban. Training a deep learning model using a dataset that depicts a bimodal distribution could potentially cause an underfitting problem. While the large-scale size of the dataset considered in this work should be sufficient to avoid underfitting problems, the tradeoff is the excessive training time needed to train a model using the whole dataset. 

\begin{figure}[t]
	\centering
	\includegraphics[width=0.45\textwidth]{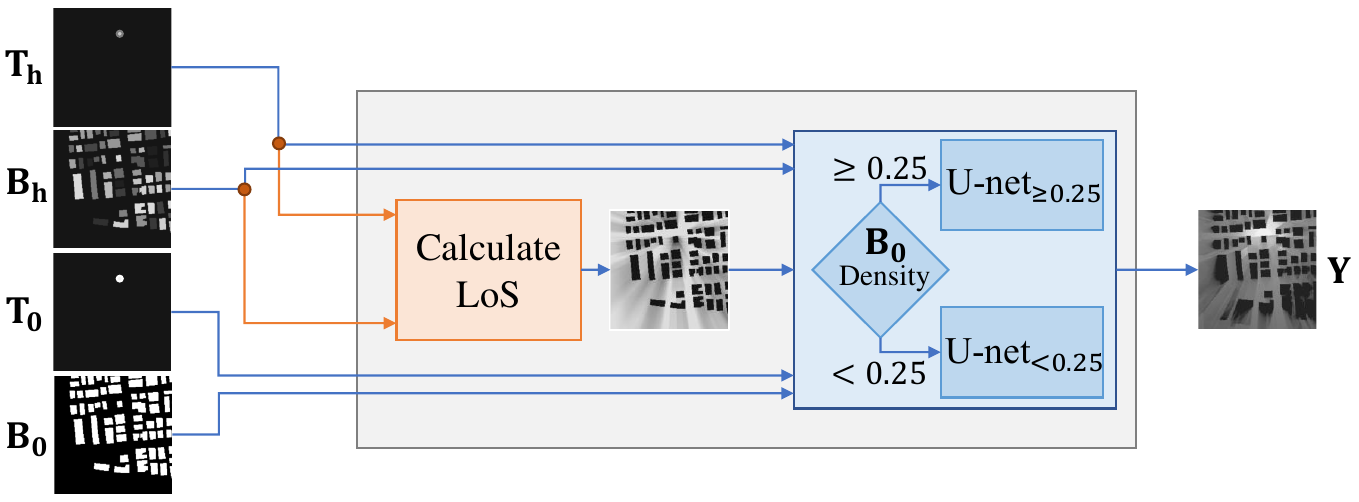}
	\caption{A visualization of the models used, where two U-nets are used during training and prediction each for a different building density. In case PxLoS$_f$ is used the two models are $\mathcal{M}(\text{noDAug}, \text{PxLoS}_f, \text{U-net}_{\geq 25}, \text{MSE})$, and $\mathcal{M}(\text{noDAug}, \text{PxLoS}_f, \text{U-net}_{< 25}, \text{MSE})$}
	\label{BdensModel}
\end{figure}

An alternative approach to address the underfitting problem, especially when using a limited dataset size, is to follow the same intuition used in modeling propagation environments and design a \textit{specialized} model for the targeted environment. To this end, we select a building density threshold of 25\% to split the dataset and train two different models. Using a building density threshold of 25\% results in 418 maps with building density $>$ 25\% ($\approx$ 60\% of the dataset) and 282 maps with building density $\leq$ 25\% ($\approx$ 40\%). In order to train two different models based on the building density, for a given set of training maps, namely, $\textbf{B}_0$, $\textbf{B}_h$, $\textbf{T}_0$, $\textbf{T}_h$, and $\textbf{L}_f$, with the corresponding output image $\textbf{Y}$, we first check the building density in $\textbf{B}_0$. As depicted in Figure \ref{BdensModel}, if it is $\geq$ 25\% we use the corresponding training sample in training a u-net model, denoted by U-net$_{\geq25}$, otherwise we use the training sample in training a u-net model denoted by U-net$_{<25}$. Similarly, in the test phase, we first check the building density in $\textbf{B}_0$, and subsequently use the corresponding u-net model for the \ac{REM} prediction. The performance evaluation of \ac{REM} prediction when using two buildings-density-dependent models follows in Section \ref{sec:REMresults}.
\subsection{Data augmentation} \label{subsec: DA} 
\begin{figure}[t]
	\centering
	\includegraphics[width=0.45\textwidth]{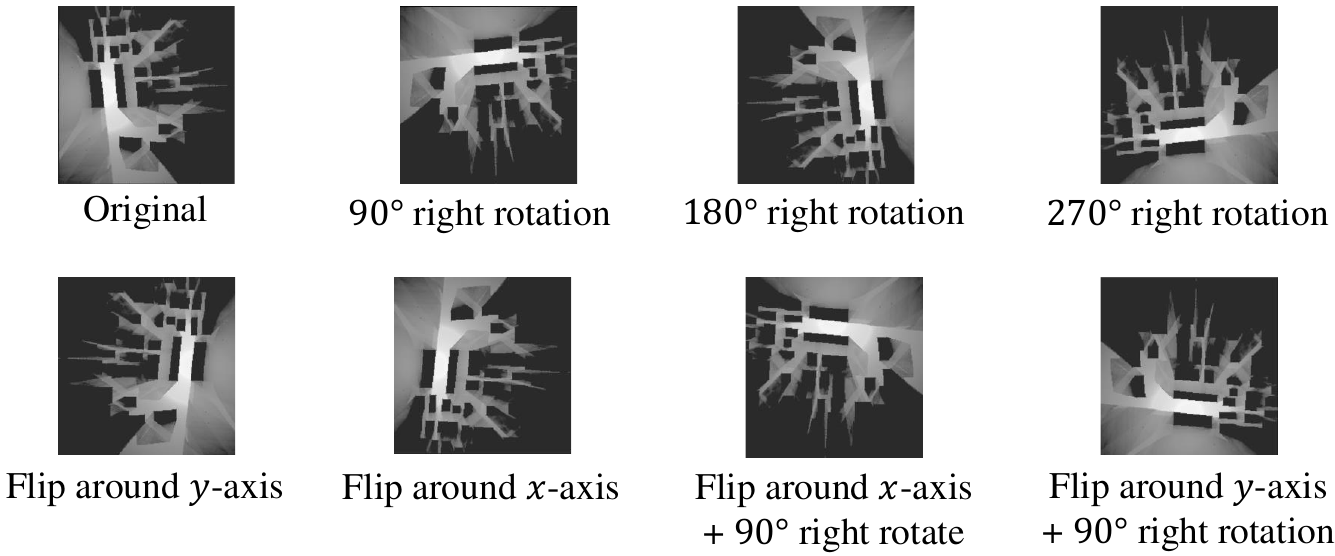}
	\caption{The considered data augmentation technique performed on a sample REM image, which in addition to the original image results in $\times$8 nonidentical images.}
	\label{Daug}
\end{figure}

\begin{figure}[t]
	\centering
	\includegraphics[width=0.45\textwidth]{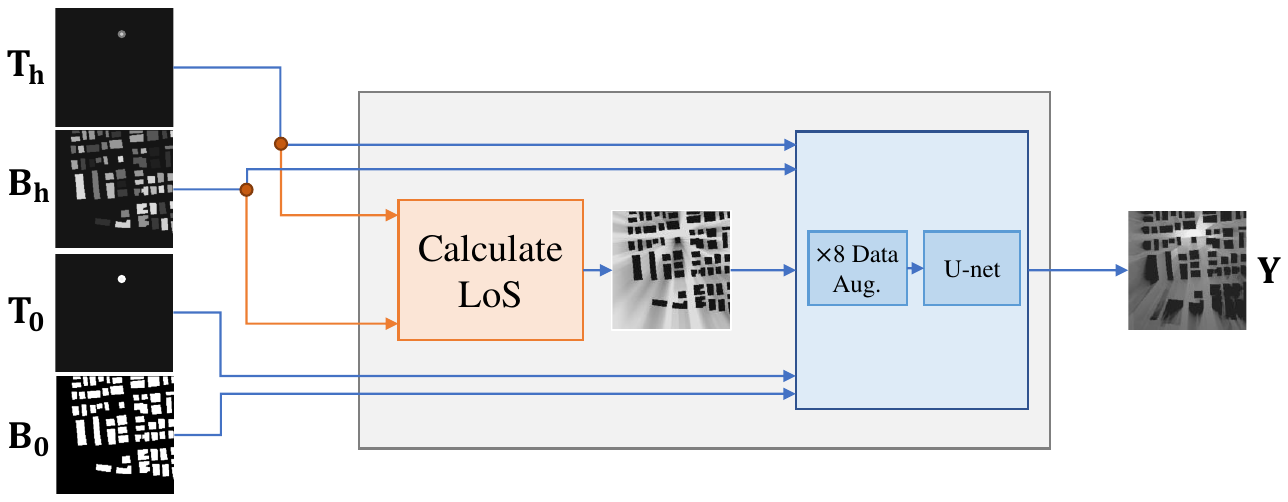}
	\caption{A visualization of the model used with a $\times 8$ data augmentation step, denoted by $\mathcal{M}(\text{DAug}, \text{PxLoS}_f, \text{U-net}, \text{MSE})$.}
	\label{DaugModel}
\end{figure}

A common practice when building deep learning models is data augmentation employed to increase the model's ability to generalize as it adds variability to the data, which in turn minimizes overfitting. This practice is particularly advantageous in \ac{REM} prediction problems as it saves on the time and cost of collecting, or using ray tracing to simulate, additional labeled data. Despite the large-scale dataset considered in this work, data augmentation showed performance gains when using horizontal, vertical, and diagonal flips \cite{lee2023pmnet}, resulting in, including original images, $\times$4 dataset size. In order to take advantage of data augmentation benefits, while maintaining a consistent 256$\times$256 pixels image size, we adopt a data argumentation technique that covers all possible rotations and flips, resulting in 8 nonidentical versions of an image, and hence a $\times$8 dataset size. In particular, for our data augmentation technique, we consider the rotations and flips illustrated in Figure \ref{Daug}. The data argumentation technique is applied during the training phase on all input images, i.e., $\textbf{B}_0$, $\textbf{B}_h$, $\textbf{T}_0$, $\textbf{T}_h$, and $\textbf{L}_f$, as well as on the corresponding output image $\textbf{Y}$, as visualized in Figure \ref{DaugModel}. We present the performance gains obtained by performing such a data argumentation technique on the dataset in the \ac{CF-mMIMO} use case in Section \ref{sec:CF}. In this use case, data argumentation plays a crucial role due to the limited size of the considered dataset.

\begin{remark}
In scenarios where we examine building-density-based data split and data augmentation, we deal with rather limited dataset size, e.g., only 100 city maps or even one in the CF-mMIMO use case (cf. Section \ref{sec:CF}). In these scenarios, we opt for u-net input supported by \textbf{B}$_0$ and \textbf{T}$_0$, as these two extra images enrich the input data and positively affect the RMSE performance. This positive impact disappears in cases where we significantly increase the size of the training data, e.g., 600 maps, and ends up being a burden that slows the training time. Therefore, in such scenarios, we only use \textbf{B}$_h$, \textbf{T}$_h$, and \textbf{L}$_f$ to construct the input. 
\end{remark}

\subsection{Alternative deep learning architecture based on \ac{KL} divergence loss} \label{subsec: KL} 
\begin{figure}[t]
	\centering
	\includegraphics[width=0.45\textwidth]{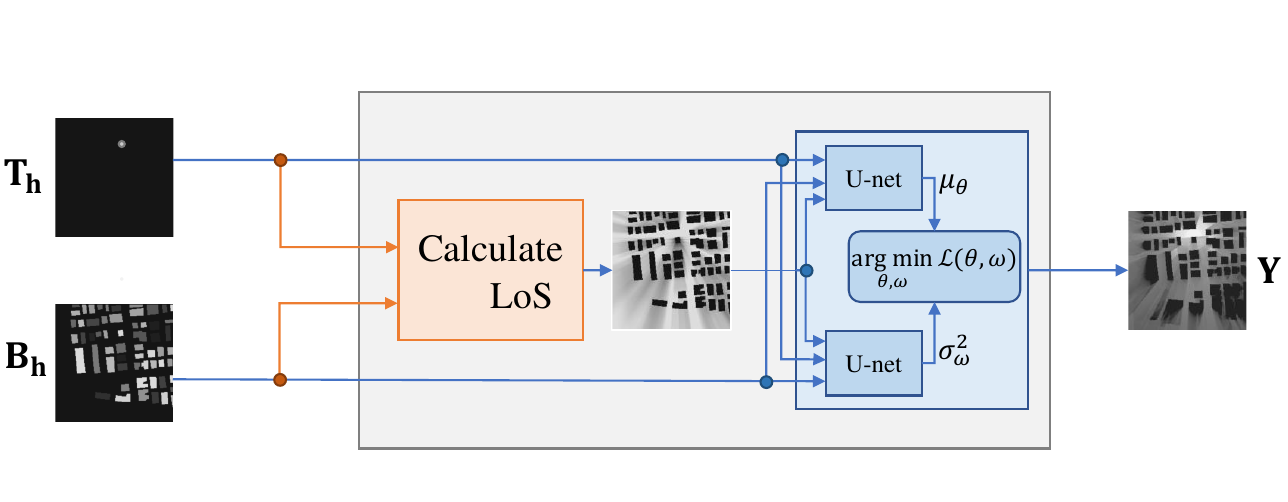}
	\caption{A visualization of the KL-based model architecture adopted in~\cite{krijestorac2021spatial}, in which two identical u-nets are used, one for mean and one for variance prediction.}
	\label{KL-Model}
\end{figure}

In this section, we present an alternative deep learning architecture for REM prediction. This architecture can be used with all the preprocessing discussed previously. In this approach, instead of using the \ac{MSE} loss function, we exploit the KL divergence loss function~\cite{kullback1951information}. Specifically, we assume that the probability distribution of signal strength follows the Gaussian distribution and train two identical u-nets, one for mean and one for variance prediction, for estimating the distribution of the signal strength, as depicted in Figure \ref{KL-Model}. We use the KL divergence as the loss function as it is well suited for measuring differences in probability distributions. Our primary reason for using this loss function is that it is known to act as an intelligent regression function: locations for which the model learned to predict high uncertainty will have a smaller effect on the loss. Such locations are often the ones with extremely low gain, where prediction is challenging for deep-learning models. After training, the u-net for estimating variance is discarded and the one for estimating mean is used for REM prediction in the test phase. We have used this approach for REM prediction, but for a different problem setup, in one of our previous works~\cite{krijestorac2021spatial}. For the sake of brevity, we do not repeat the details here, but interested readers are encouraged to refer to~\cite{krijestorac2021spatial} for more details. The architecture of the u-nets used in this approach is the same as the one used in~\cite{krijestorac2021spatial}, with the only difference being the size of the radio maps: 256$\times$256 in this paper and 64$\times$64 in \cite{krijestorac2021spatial}.

\textit{U-net models notation}: Considering the different LoS preprocessing methods, data augmentation, and environment building density, we use $\mathcal{M}(a,b,c,d)$ to denote the different u-net-based models we explore in this work, where 
\begin{itemize} 
	\item $a \in \{\text{DAug, noDAug }\}$ corresponds to data augmentation and no data augmentation.
	\item $b \in \{ \text{noLoS}, \text{PxLoS}_f, \text{AbLoS}_f, \text{NNLoS}_f\}$ indicates the type of LoS calculation used.
	\item $c \in \{ \text{U-net}, \text{U-net}_{\geq 25}, \text{U-net}_{< 25} \}$
	\item $d \in \{\text{MSE}, \text{KL} \}$
\end{itemize}

\section{Evaluation Results of Rem Prediction} \label{sec:REMresults}

This section details the evaluation results of our \ac{REM} prediction approach, quantifying the impact of our key design parameters, namely, \ac{LoS} maps, building-density-based training, and the loss function. In this section, we used \textit{RadioMap3DSeer} dataset maps 0-500, 500-600, and 600-700 for model training, validation, and testing, respectively, unless otherwise mentioned. At the end of this section, we also present the results we submitted to the ICASSP challenge test set.

\subsection{Performance Metric}

Considering the output REM $\textbf{Y}$ and the corresponding estimated REM $\hat{\textbf{Y}}$, we define the \ac{RMSE} of a test set containing $L$ maps as
\begin{equation}
	\mathcal{E}_{\text{RMSE}} = \sqrt{ \frac{\sum_L \sum_N |\textbf{Y} - \hat{\textbf{Y}} |^2}{LN}} \,,
	\label{RMSE_t}
\end{equation} 
where $N$ is the total number of pixels, which equals to $256\times256$. Note that this RMSE slightly differs from the RMSE used in the \textit{2023 IEEE ICASSP First Pathloss Radio Map Prediction Challenge}~\cite{krijestorac2023agile,lee2023pmnet}, where building locations were set to zero before calculating the \ac{RMSE} of a given map. By setting buildings locations to zero, one corrects any estimation error in the building pixels of the estimated REM, slightly underestimating the RMSE error compared to the RMSE in (\ref{RMSE_t}).

\begin{figure}[t]
	\centering
%
%
\definecolor{mycolor1}{rgb}{0.00000,0.44700,0.74100}%
\definecolor{mycolor2}{rgb}{0.85000,0.32500,0.09800}%
\definecolor{mycolor3}{rgb}{0.92900,0.69400,0.12500}%
\definecolor{mycolor4}{rgb}{0.46667,0.67451,0.18824}%

\begin{tikzpicture}
\begin{axis}[
width=8cm,
height=6cm,
xmin=0.5,
xmax=4.5,
xtick={1,2,3,4},
xlabel={LoS preprocessing approach},
xticklabels={No LoS map,PxLoS,AbLoS,NNLoS},
ytick={0.03,0.035,0.04},
ymin=0.03,
ymax=0.04,
ylabel={Normalized RMSE},
xmajorgrids,
ymajorgrids,
y tick label style={
	/pgf/number format/.cd,
	fixed,
	fixed zerofill,
	precision=3,
	/tikz/.cd
},
every axis plot/.append style={
	ybar,
	bar width=.4,
	bar shift=0pt,
	draw=black,
	fill
}
]
	\addplot[mycolor1,draw=black]coordinates {(1,0.039133977)};
	\addplot[mycolor2,draw=black]coordinates{(2,0.0342862)};
	\addplot[mycolor3,draw=black]coordinates{(3,0.03577989)};
	\addplot[mycolor4,draw=black]coordinates{(4,0.03696104)};
\end{axis}
\end{tikzpicture}
	\caption{Impact of different LoS preprocessing approaches on REM prediction accuracy.}
	\label{fig:los_methods}
\end{figure}
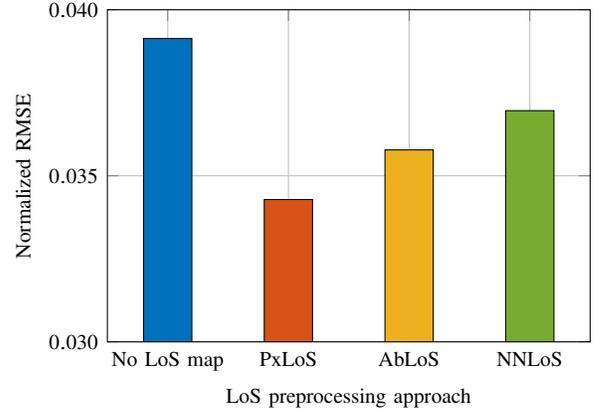
\subsection{Impact of LoS maps} \label{subsection:los_results}
In Figure \ref{fig:los_methods}, we use the u-net architecture of Figure~\ref{fig:unet_arch} for REM prediction. For the bar labeled `No LoS map,' no LoS map is used, and only $\textbf{T}_h$ and $\textbf{B}_h$ are used as u-net input, i.e., $K=2$. For the remaining three bars, we use $\textbf{T}_h$,  $\textbf{B}_h$, and $\textbf{L}_f$ as u-net input, i.e., $K=3$. The labels on X-axis in Figure~\ref{fig:los_methods} indicate the method for computing the LoS map, $\textbf{L}_f$. We make several observations for this figure. 

First, by comparing the REM prediction accuracy for the `No LoS map' based u-net and PxLoS-based u-net, we see that including the LoS map significantly improves the prediction accuracy; specifically, the prediction accuracy improves by 13\%. As discussed in Section~\ref{sec:los_compute_methods}, this happens because using the LoS map, based on domain knowledge of RF propagation, helps the u-net to learn a better mapping easily. 

Second, the AbLoS-based u-net and NNLoS-based u-net also perform better than the `No LoS map' based u-net. This is expected due to using LoS maps in AbLoS-based u-net and NNLoS-based u-net. However, we observe that the REM prediction accuracy with AbLoS-based u-net and NNLoS-based u-net is lower than that with PxLoS-based u-net. The reason is that both AbLoS and NNLoS use more approximations than PxLoS when computing the LoS map. As explained in Section~\ref{sec:los_compute_methods}, NNLoS tries to mimic PxLoS via a neural network, and AbLoS assumes all the pixels are equidistant from the transmitter along the driving dimension. 

\begin{figure}[t]
	\centering
\includegraphics[width=0.45\textwidth]{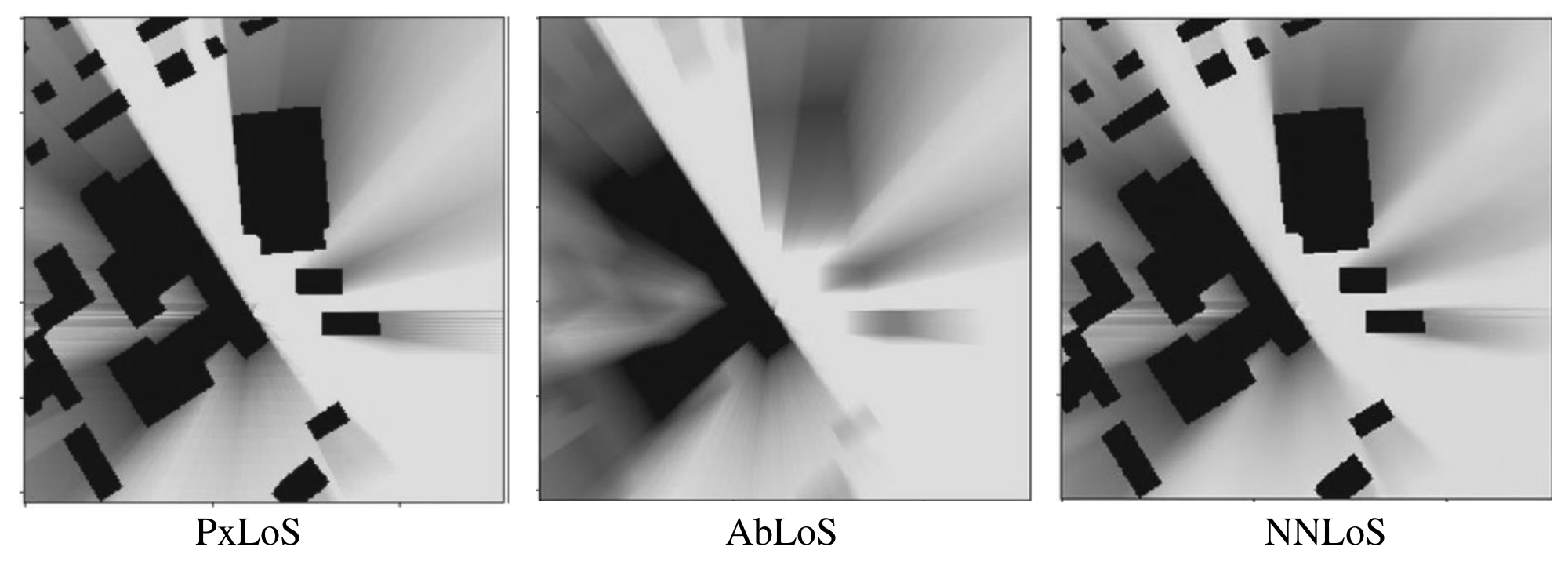}
	\caption{Visual comparison of LoS maps generated by different methods.}
	\label{fig:los_examples}
\end{figure}

Third, we see that AbLoS-based u-net performs slightly better than NNLoS-based u-net. To better understand the reason behind that, we also plot the generated LoS maps using three methods for one scenario in Figure~\ref{fig:los_examples}. This figure shows that the LoS map predicted by NNLoS is very similar to the one computed by the PxLoS method, and the LoS map created by AbLoS looks more different than the other two. Hence, we expect NNLoS-based u-net to have better accuracy than AbLoS-based u-net. However, that is not the case in Figure~\ref{fig:los_methods}. We believe the reason for this discrepancy is the following. 

In NNLoS-based u-net, first, we train the LoS predictor u-net using all the available training examples. Then when the REM predictor u-net is trained, we use the LoS maps predicted by NNLoS for all the training examples. I.e., the predicted LoS maps are for the same set of examples that were used for the training of NNLoS. This way, the REM predictor u-net is trained on LoS maps that are biased. This limits the generalization capability of the NNLoS-based u-net and adversely impacts the REM prediction accuracy on unseen test data.

Next, we compare the different methods for \ac{LoS} map generation in terms of overall prediction time in Table.~\ref{tab:pred_time}. Here, overall prediction time is the sum of the time required for generating the LoS map and predicting the REM. Although PxLoS-based u-net has the highest REM prediction accuracy, we see from Table.~\ref{tab:pred_time} that its overall prediction time is the maximum. Both AbLoS-based u-net and NNLoS-based u-net are significantly faster than PxLoS-based u-net, but their overall prediction time is almost double the prediction time of the `No LoS map' based u-net. `No LoS map' based u-net has the lowest overall prediction time because it performs no preprocessing, only forward pass of the u-net. Lastly, we see that both AbLoS-based u-net and NNLoS-based u-net have comparable overall prediction times.

\begin{table}[t!]
\caption{Average prediction time per REM. This includes both preprocessing and forward pass of the neural network.}
\label{tab:pred_time}
\centering{
\resizebox{3in}{!}
{
\begin{tabular}{| c |c |c| c |}
\hline
\textbf{No LoS map} & \textbf{PxLoS} & \textbf{AbLoS} & \textbf{NNLoS}\\
 \hline
 \hline
6.5 msec & $3 \times 10^3$ msec & 14 msec & 13 msec \\
\hline
\end{tabular}}}
\end{table}

\begin{table}[t]
	\caption{Summary of the results we presented in the challenge using a new unpublished test set from the organizers.}
	\centering
	\begin{tabular}{ | c || c | c |}
		\hline
		\scriptsize{\textbf{Model}} & \scriptsize{\textbf{Norm. RMSE}} & \scriptsize{\textbf{Run Time [ms]}}\\
		\hline
		\hline
		$\mathcal{M}(\text{noDAug}, \text{noLoS}, \text{U-net}, \text{MSE})$ & 0.051 & 6 \\ 
		\hline
		$\mathcal{M}(\text{noDAug}, \text{AbLoS}_f, \text{U-net}, \text{MSE})$  
		& 0.046 & 14\\ 
		\hline
		$\mathcal{M}(\text{noDAug}, \text{AbLoS}_f, \text{U-net}, \text{KL})$ & 0.045 & 14 \\
		\hline
	\end{tabular}
	\label{resultsSum}
\end{table}

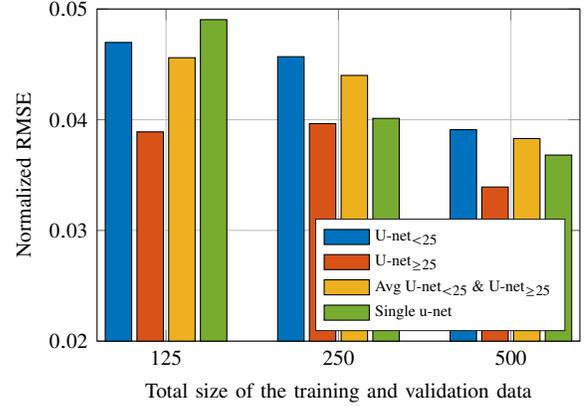
\begin{figure}[t]
	\centering
%
%
\definecolor{mycolor1}{rgb}{0.00000,0.44700,0.74100}%
\definecolor{mycolor2}{rgb}{0.85000,0.32500,0.09800}%
\definecolor{mycolor3}{rgb}{0.92900,0.69400,0.12500}%
\definecolor{mycolor4}{rgb}{0.46667,0.67451,0.18824}%
\begin{tikzpicture}

\begin{axis}[%
legend style={font=\fontsize{6}{5}\selectfont},
width=8cm,
height=6cm,
bar shift auto,
symbolic x coords={125,250,500},
enlarge x limits=0.2,
xtick=data,
xlabel={Total size of the training and validation data},
ymin=0.02,
ymax=0.05,
ylabel={Normalized RMSE},
axis background/.style={fill=white},
xmajorgrids,
ymajorgrids,
legend style={at={(0.97,0.03)}, anchor=south east, legend cell align=left, align=left, draw=white!15!black}
]
\addplot[ybar, fill=mycolor1, draw=black, area legend] table[row sep=crcr] {%
125	0.04698\\
250	0.045692205\\
500	0.0391\\
};
\addplot[forget plot, color=white!15!black] table[row sep=crcr] {%
125	0\\
500	0\\
};
\addlegendentry{$\text{U-net}_{{<25}}$}

\addplot[ybar, fill=mycolor2, draw=black, area legend] table[row sep=crcr] {%
125	0.0389\\
250	0.03965\\
500	0.03392\\
};
\addplot[forget plot, color=white!15!black] table[row sep=crcr] {%
125	0\\
500	0\\
};
\addlegendentry{$\text{U-net}_{{\geq25}}$}

\addplot[ybar, fill=mycolor3, draw=black, area legend] table[row sep=crcr] {%
125	0.0456\\
250	0.044\\
500	0.0383\\
};
\addplot[forget plot, color=white!15!black] table[row sep=crcr] {%
125	0\\
500	0\\
};
\addlegendentry{$\text{Avg U-net}_{<25}\text{ \& U-net}_{{\geq25}}$} 

\addplot[ybar, fill=mycolor4, draw=black, area legend] table[row sep=crcr] {%
125	0.049038\\
250	0.04012\\
500	0.036806\\
};
\addplot[forget plot, color=white!15!black] table[row sep=crcr] {%
125	0\\
500	0\\
};
\addlegendentry{Single u-net}

\end{axis}
\end{tikzpicture}%
	\caption{Impact of using two separate models based on the building density compared to the case of using a single model. Training-validation split is 80-20 \%.}
	\label{BdnstyRMSE}
\end{figure}

\subsection{Building density}
Going beyond the challenge-required results, we also consider scenarios with constraints on the size of training data and training time. Despite the nearly instant prediction time of our proposed model, which takes only a few milliseconds per map, the training time with 600 $\times$ 80 maps the whole dataset takes tens of hours, depending on the available computational resource, i.e., the number of GPU clusters. In Figure \ref{BdnstyRMSE}, we present the normalized RMSE performance when training two different u-net models, $\mathcal{M}(\text{noDAug},\text{PxLoS}_f, \text{U-net}_{\geq 25}, \text{MSE})$ and $\mathcal{M}(\text{noDAug},\text{PxLoS}_f, \text{U-net}_{< 25}, \text{MSE})$, based on the map's building density, compared to the case in which a single u-net $\mathcal{M}(\text{noDAug},\text{PxLoS}_f, \text{U-net})$ is used. This figure investigates scenarios with constraints on the training time or the size of the training data due to limitations on training computational time or resources. The figure considers three sizes of training data, 125, 250, and 500, all split 80\% for training and 20\% for validation. In all the models shown in Figure \ref{BdnstyRMSE}, maps 600-700 from the \textit{RadioMap3DSeer} are used as a test set. In the case of $\mathcal{M}(\text{noDAug},\text{PxLoS}_f, \text{U-net}_{\geq 25}, \text{MSE})$ and $\mathcal{M}(\text{noDAug},\text{PxLoS}_f, \text{U-net}_{< 25}, \text{MSE})$ models, the dataset is approximately 40\% and 60\%, respectively, both for training and validation. However, the test set is biased with 84\% used the model $\mathcal{M}(\text{noDAug},\text{PxLoS}_f, \text{U-net}_{< 25}, \text{MSE})$ and 16\% used the model $\mathcal{M}(\text{noDAug},\text{PxLoS}_f, \text{U-net}_{\geq 25}, \text{MSE})$. Note that we used this test set despite its bias to ensure consistency across the paper. The training and test process follows the block diagram shown in Figure \ref{BdensModel}. As shown in the figure, in the case of a limited dataset, e.g., 125 maps, training a model based on the building density in the area of interest outperforms the case in which a single model is used without paying attenuation to the building density. However, increasing the size of the training data gives an edge performance of 0.004 in the case where a single model is used compared to the case where building-density-based models are used. This difference in performance diminishes to only 0.001 in the case where the size of the training dataset is 500, indicating that all models considered in the figure have sufficient data to generalize and avoid overfitting.

\subsection{Impact of KL divergence loss}
\begin{figure}[t]
	\centering
%
%
\definecolor{mycolor1}{rgb}{0.00000,0.44700,0.74100}%
\definecolor{mycolor2}{rgb}{0.85000,0.32500,0.09800}

\begin{tikzpicture}
	\begin{axis}[
		width=8cm,
		height=6cm,
		xmin=0.5,
		xmax=2.5,
		xtick={1,2},
		xlabel={Loss function used},
		xticklabels={MSE loss, KL loss},
		ytick={0.03,0.035,0.04},
		ymin=0.03,
		ymax=0.04,
		ylabel={Normalized RMSE},
		xmajorgrids,
		ymajorgrids,
		y tick label style={
			/pgf/number format/.cd,
			fixed,
			fixed zerofill,
			precision=3,
			/tikz/.cd
		},
		every axis plot/.append style={
			ybar,
			bar width=.4,
			bar shift=0pt,
			draw=black,
			fill
		}
		]
		\addplot[mycolor1,draw=black]coordinates {(1,0.03577989)};
		\addplot[mycolor2,draw=black]coordinates{(2,0.03442600288076731)};
	\end{axis}
\end{tikzpicture}
	\caption{Impact of loss function on the REM prediction.}
	\label{fig:compare_loss}
\end{figure}
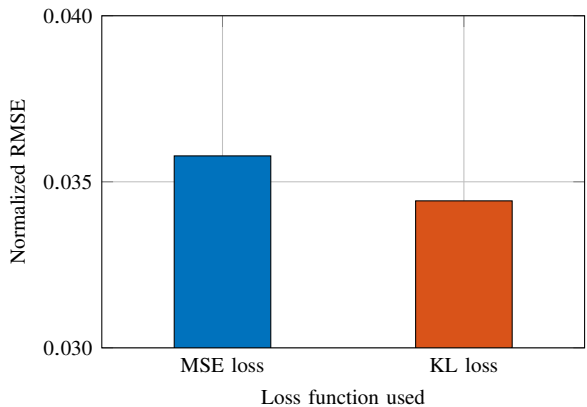

In Figure~\ref{fig:compare_loss}, we show the impact of the loss function on REM prediction. Specifically, for the bar labeled `MSE,' we used the u-net of Figure~\ref{fig:unet_arch} and trained it using the mean squared error (MSE) loss.  For the bar labeled `KL,' we used the u-net architecture and loss function as discussed in Section~\ref{subsec: KL}. For both of these approaches, we use 
$\textbf{T}_h$, $\textbf{B}_h$, and $\textbf{L}_f$ as input, and $\textbf{L}_f$ is computed using the AbLoS method. We observe from this figure that using the KL divergence loss provides some improvement in REM prediction accuracy over the MSE loss.

\subsection{Challenge Results}

The proposed models are evaluated on a new unpublished test set from the challenge organizers, consisting of 84 city maps, each with 80 different transmitter locations, resulting in a total of 6720 radio maps. Similar to the \textit{RadioMap3DSeer} dataset, receivers are assumed to have a fixed height of 1.5\,m, and transmitters are assumed to be located 3\,m above buildings' rooftops. Table \ref{resultsSum} summarizes the results obtained using our proposed models. As shown in the table, $\mathcal{M}(\text{noDAug}, \text{AbLoS}_f, \text{U-net}, \text{KL})$ outperforms $\mathcal{M}(\text{noDAug}, \text{AbLoS}_f, \text{U-net}, \text{MSE})$ and $\mathcal{M}(\text{noDAug}, \text{noLoS}, \text{U-net}, \text{MSE})$, with 8\,ms extra runtime for \ac{LoS} calculation as a tradeoff compared to $\mathcal{M}(\text{noDAug}, \text{noLoS}, \text{U-net}, \text{MSE})$.

\section{Use-Case: AP ON/OFF Switching in CF-mMIMO} \label{sec:CF}
\Ac{CF-mMIMO} is a promising novel wireless network architecture proposed to address the inadequate cell-edge users performance, which might experience tens of dB weaker channel compared to cell-center users \cite{demir2021foundations}. Instead of using a single \ac{AP} to serve a \ac{UE}, in \ac{CF-mMIMO} networks, a \ac{UE} is alternatively served by several, or even all, \acp{AP} \cite{interdonato2019ubiquitous}. In \ac{CF-mMIMO} networks, distributed \acp{AP} jointly operate to coherently serve \acp{UE} on the same time/frequency resource, using spatial multiplexing. This joint operation of distributed \acp{AP} ensures not only a higher \ac{SNR}, but also better multi-user interference suppression when compared to the conventional cellular networks \cite{demir2021foundations}. However, in order to harness the advantages of \ac{CF-mMIMO} networks, a dense deployment with large number of \acp{AP} is needed. While the \acp{AP} used in \ac{CF-mMIMO} are expected to be significantly less complex than conventional base stations used in cellular networks, the large-scale deployment needed raises concerns about the overall power consumption and the corresponding energy-related pollution \cite{chen2022survey}.

In order to address the power-consumption concerns in \ac{CF-mMIMO} networks, strategies known as \ac{ASO} are attracting considerable research focus \cite{chen2022survey}. \ac{ASO} strategies suggest that \acp{AP} should be dynamically switched on (put in active mode) and off (put in sleep mode) based on the \acp{UE} traffic demand. In particular, \ac{ASO} strategies consider the \ac{AP} status as an optimization variable, aiming at selecting a subset of \acp{AP} that meets \acp{UE} \ac{SE} requirements while the remaining \acp{AP} are switched off. Selecting the optimal energy-efficient subset of \acp{AP} that meets a given single or multiple \acp{UE} \ac{SE} requirements is an NP-hard problem, requiring assessing all possible combinations of \acp{AP} \cite{van2020joint}. The authors in \cite{van2020joint} presented a global solution of the \ac{AP} selection problem by solving a computationally-intensive non-convex optimization problem. To address the computational complexity of the non-convex optimization problem, heuristic \ac{ASO} strategies based on the location and propagation losses between \acp{AP} and \acp{UE} are introduced in \cite{femenias2020access,garcia2020energy,van2020joint}, including \ac{ASO} strategies such as random selection ASO (RS-ASO), nearest neighbor ASO, Chi-square test-based ASO (ChiS-ASO), optimal energy-efficiency-based greedy ASO, and \ac{MPL-ASO}. Among these strategies, \ac{MPL-ASO} reportedly provided a good tradeoff between the \ac{SE} performance and complexity, exploiting large-scale fading coefficients between APs and UEs with a minor performance penalty \cite{femenias2020access,chen2022survey}.

ASO strategies that depend on propagation losses assume that \acp{AP} are frequently turned on to collect the measurements needed to select the optimal set of \acp{AP}, which implies that some \acp{AP} might wake-up to do channel measurements and end up not serving any \acp{UE}, negatively affecting the energy efficiency of the whole network. This particular problem, combined with the promising performance of \ac{MPL-ASO}, motivates our work on the \acp{ASO} use-case. In this section, we present a use-case demonstrating that one can exploit the swift and accurate \ac{REM} prediction to boost the energy efficiency of \ac{MPL-ASO} by eliminating the need to unnecessarily turn on \acp{AP} to do channel measurements. By using the subset of active \acp{AP}, we train our u-net-based model to predict the large-scale fading of the off \acp{AP}, and subsequently use the predicted large-scale fading to decide on which extra \acp{AP} to switch on. In the following, we detail the system model and the evaluation results.

\begin{figure}[t]
	\centering
	\includegraphics[width=0.45\textwidth]{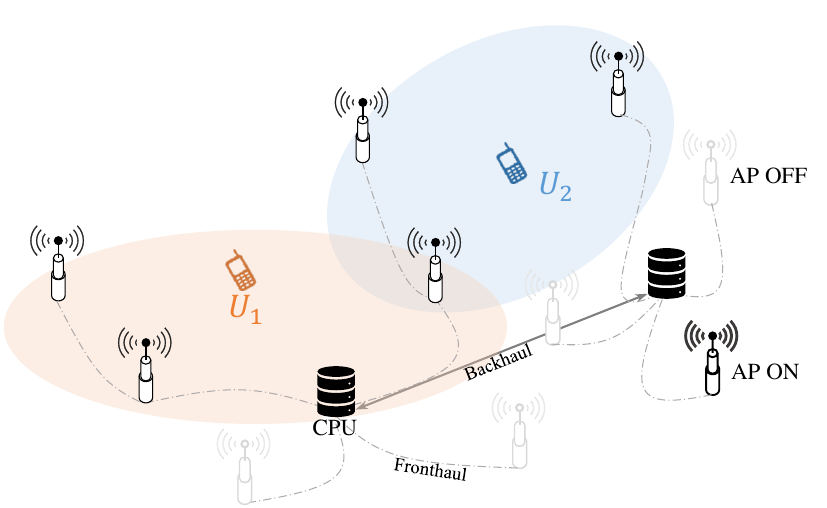}
	\caption{A representation of a CFmMIMO network with ASO.}
	\label{CF-mMIMO_model}
\end{figure}

\subsection{System Model}
We consider a \ac{CF-mMIMO} network with $M$ single-antenna \acp{AP} deployed in an outdoor environment, serving $N_u$ arbitrarily distributed users in an outdoor area of interest. \acp{AP} are connected via the so-called fronthaul connections to one or multiple edge-cloud \acp{CPU}, which are linked via an optical fiber or a microwave backhaul link, as depicted in Figure \ref{CF-mMIMO_model}. We assume that an \ac{MPL-ASO} strategy is implemented in the network, which implies that each of the \acp{AP} can be either in active/ON mode or sleep/OFF mode, depending on \ac{LSF} fading between the \acp{AP} and \acp{UE}. At any given time instant, we have a set of active \acp{AP}, $\mathcal{A}_a = \{A_1, \dots, A_{M_a}\}$ and a set of sleep \acp{AP}, $\mathcal{A}_s = \{A_1, \dots, A_{M_s}\}$, with $M_a + M_s = M$, and $M_a \neq 0$. In case a connected \ac{UE} demands extra downlink \ac{SE}, the \ac{CPU} coordinates the \ac{MPL-ASO} strategy, deciding on the number and the addresses of \acp{AP} to be switched on. It is worth noting that in this use-case, we mainly focus on the performance of \ac{MPL-ASO} with predicted \ac{LSF} path gains; for a detailed formulation of the downlink \ac{SE}, we refer the reader to Chapter 6 in \cite{demir2021foundations}.

\subsection{LSF-Prediction-Based MPL-ASO}
Our main objective is to exploit \ac{LSF} to decide on which extra \acp{AP} to switch on in order to join the already active \acp{AP} in serving the extra \ac{SE} demanding \ac{UE}. In other words, we aim at finding a subset of sleep \acp{AP} to reactivate, $ \mathcal{A}_{sa} = \{A_1, \dots, A_{M_{sa}}\} \subseteq \mathcal{A}_s$, with $M_{sa} \leq M_s$. To this end, we train a model $\mathcal{M}$ using the \ac{LSF} information, which is assumed to be collected from the set of active \acp{AP} $\mathcal{A}_a$ and outdoor users they served throughout their active time window. We explore the case in which the full \ac{REM} of active APs is available, as well as the case where only \ac{LSF} information from a set of randomly distributed users is available. Using the obtained $\mathcal{M}$ model, we stack the predictions of all \acp{AP} $\in \mathcal{A}_{s}$ in $\hat{\textbf{Y}}_u$, and subsequently, order them based on \ac{LSF} to the \ac{UE} of interest from lowest to highest. We orderly activate \acp{AP} from the set $\mathcal{A}_{s}$, i.e., add them to the estimated subset $\hat{\mathcal{A}}_{sa}$ until the \ac{UE}'s \ac{SE} is satisfied. We summarized the steps of the \ac{LSF}-prediction-based \ac{MPL-ASO} in Algorithm \ref{alg:MPL-ASO}. In the following, we present the evaluation results concerning our \ac{LSF}-prediction-based \ac{MPL-ASO} use case.

\begin{algorithm} [t]
	\caption{LSF Prediction-based MPL-ASO}
	\label{alg:MPL-ASO}
	\begin{algorithmic}[1]
		\STATE \textbf{Input:} $\textbf{B}_h$, $\textbf{B}_0$, $\textbf{T}_h$, $\textbf{T}_0$, $\mathcal{A}_{a}$, UE's location and SE demand.
		\STATE \textbf{Output:} $\hat{\mathcal{A}}_{sa}$
		\item[]
		\STATE Train prediction model $\mathcal{M}$ using $\mathcal{A}_{a}$
		\FOR{each extra SE demanding UE}
		\STATE $\hat{\textbf{Y}}_u$ $\leftarrow$ $\mathcal{M}(\text{DAug}, \text{PxLoS}_f, \text{U-net}, \text{MSE}) \,\, \forall \, \mathcal{A}_{s}$
		\STATE Order \acp{AP} in $\mathcal{A}_{s}$ based on $\hat{\textbf{Y}}_u$
		\STATE $\hat{\mathcal{A}}_{sa}$ $\leftarrow$ Orderly activate \acp{AP} from $\mathcal{A}_{s}$ until SE is satisfied
		\ENDFOR
		\RETURN $\hat{\mathcal{A}}_{sa}$
	\end{algorithmic}
\end{algorithm}

\begin{figure}[t]
	\centering
	\includegraphics[width=0.45\textwidth]{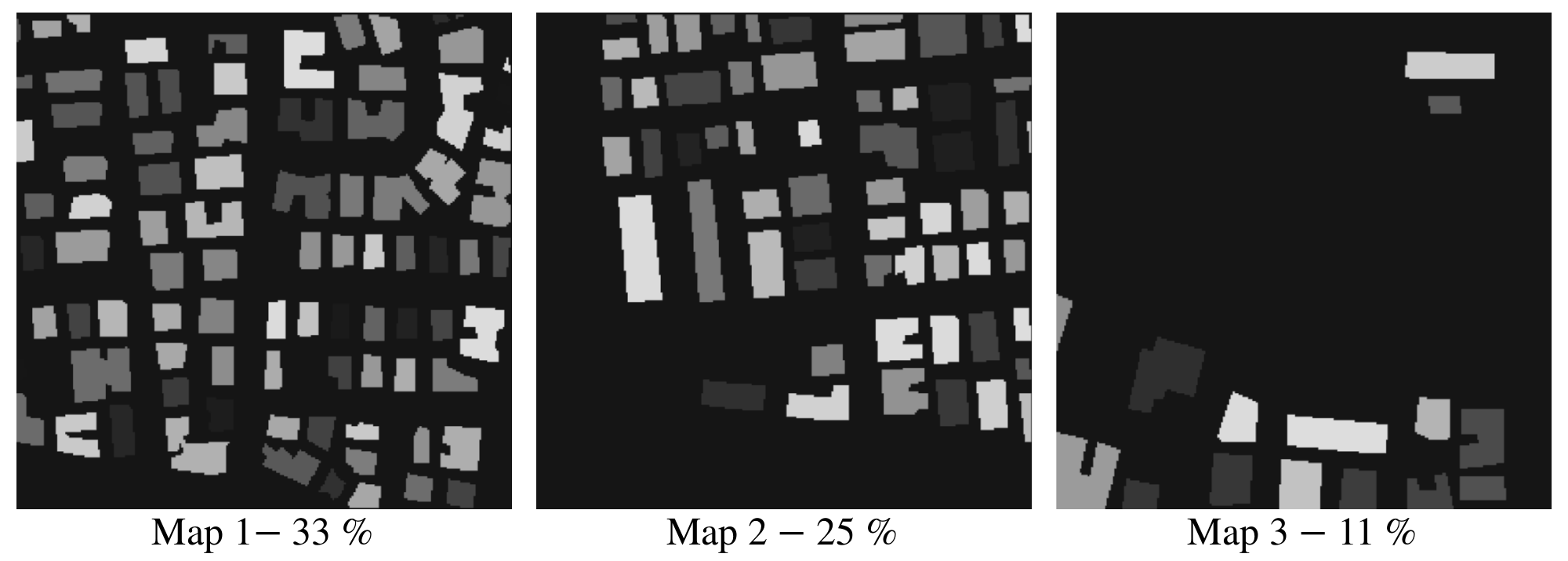}
	\caption{Maps layout considered in the ASO-MPL use case, with building density from left to right of 33\%, 25\%, and 11\%.}
	\label{ASOmaps}
\end{figure}

\subsection{Evaluation Results}
In this section, we present our results on \ac{MPL-ASO} \ac{AP} selection using predicted \ac{LSF} based on our REM model. In order to evaluate the AP selection performance in various propagation environments, we consider three city maps, namely, map 1, map 2, and map 3 with building density of 33\%, 25\%, and 11\%, respectively, as illustrated in Figure \ref{ASOmaps}. We assume that the 80 transmitter locations provided for each map in the \textit{RadioMap3DSeer} dataset represent a \ac{CF-mMIMO} network with 80 \acp{AP}. To evaluate the performance of \ac{MPL-ASO} with predicted \ac{LSF}, we used a subset of the 80 APs, e.g., 70, to represent the set of active APs $\mathcal{A}_a$, which we use to train our u-net model $\mathcal{M}(\text{DAug},\text{PxLoS}_f, \text{U-net}, \text{MSE})$, unless otherwise mentioned. Subsequently, we use this model to predict the \ac{LSF} of the remaining 10 APs. We consider training and prediction using the full \ac{REM} as well as a scattered \ac{LSF} of selected outdoor locations. The cases where the number of APs to switch on is one, two, or three. In the following, we first define our performance metrics and then present our evaluation results.

\subsubsection{Performance Metric}
In this section, we use \ac{RMSE} as a performance metric for the full \ac{REM} prediction as well as for the scattered \ac{LSF} prediction. However, unlike Section \ref{sec:REMresults}, where we calculate the \ac{RMSE} of the whole output map, in this section, we only consider the selected outdoor locations in the \ac{RMSE} calculation. The reason behind using a different \ac{RMSE} calculation here is that in this section, we present cases where only a few outdoor locations are considered. In such cases calculating RMSE over the whole map might be misleading. Considering the output map $\textbf{Y}_u$, we define the \ac{RMSE} of an estimated map $\hat{\textbf{Y}}_u$ as
\begin{equation}
	\mathcal{E}_{\text{RMSE}, {u}} = \sqrt{ \frac{\sum_L \sum_N |\textbf{Y}_u - \hat{\textbf{Y}}_u |^2}{LN_u}} \,,
	\label{RMSEu}
\end{equation} 
where $N_u$ is the number of the considered outdoor locations per map, which in the training phase represents the number of UEs, and $L$ is the total number of maps used in the test set. To evaluate the AP selection performance, we conduct AP selection for the available outdoor user locations and calculate the error percentage across these locations as follows 
\begin{equation}
	\mathcal{E}_{\text{APsel}_{u}} = \frac{\sum_{n = 1}^{N_u} \xi_n}{N_u} \,,
	\label{RMSEu}
\end{equation}
where
\begin{equation}
	\xi_n =
	\begin{cases}
		0 & \text{if $\mathcal{A}_{sa}$ = $\hat{\mathcal{A}}_{sa}$ at location $n$}\\
		1 & \text{otherwise}
	\end{cases}       
\end{equation}
with ${\mathcal{A}}_{sa}$ and $\hat{\mathcal{A}}_{sa}$ being the true and estimated set of APs to switch on.

\subsubsection{Impact of Data Augmentation}
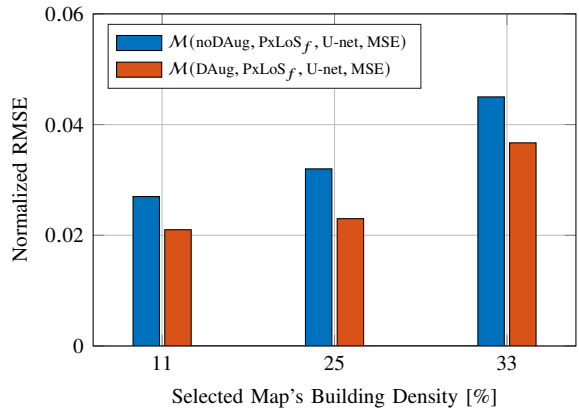
\begin{figure}[t!]
	\centering
%
%
\definecolor{mycolor1}{rgb}{0.00000,0.44700,0.74100}%
\definecolor{mycolor2}{rgb}{0.85000,0.32500,0.09800}%
\begin{tikzpicture}
\begin{axis}[%
legend style={font=\fontsize{6}{5}\selectfont},
width=8cm,
height=6cm,
bar shift auto,
symbolic x coords={11,25,33},
enlarge x limits=0.2,
xtick=data,
xlabel={Selected Map's Building Density [\%]},
ymin=0,
ymax=0.06,
ylabel={Normalized RMSE},
xmajorgrids,
ymajorgrids,
log ticks with fixed point,
legend style={at={(0.03,0.97)}, anchor=north west, legend cell align=left, align=left, draw=white!15!black}
]
\addplot[ybar, fill=mycolor1, draw=black, area legend] table[row sep=crcr] {%
33	0.045\\
25	0.032\\
11	0.027\\
};
\addplot[forget plot, color=white!15!black] table[row sep=crcr] {%
11	0\\
33	0\\
};
\addlegendentry{$\mathcal{M}(\text{noDAug},\text{PxLoS}_f, \text{U-net}, \text{MSE})$}

\addplot[ybar, fill=mycolor2, draw=black, area legend] table[row sep=crcr] {%
33	0.0367\\
25	0.023\\
11	0.021\\
};
\addplot[forget plot, color=white!15!black] table[row sep=crcr] {%
11	0\\
33	0\\
};
\addlegendentry{$\mathcal{M}(\text{DAug},\text{PxLoS}_f, \text{U-net}, \text{MSE})$}

\end{axis}

\end{tikzpicture}%
	\caption{Normalized RMSE of REM prediction of 10 off APs when training on a single map with 70 active APs, providing training data. Data augmentation is used to enrich the training data.}
	\label{1mapAug}
\end{figure}

In the \ac{MPL-ASO} use case, we perform model training on a single map with various AP locations, resulting in a rather limited training set size compared to the results presented in Section~\ref{sec:REMresults}. In order to enrich the training dataset and the model's ability to generalize, we employ the data augmentation method presented in Section~\ref{subsec: DA} to obtain $\times$8 training dataset size. Figure \ref{1mapAug} presents the $\mathcal{E}_{\text{RMSE}, {u}}$ of REM predicted for 10 off APs with 70 on APs used for model training (60 train and 10 validation). As shown in the figure, data augmentation outperforms the case without data augmentation for all three maps considered for the  \ac{MPL-ASO} use case. In particular, a gain in the $\mathcal{E}_{\text{RMSE}, {u}}$ of 0.009 for map 1 and map 2, and 0.006 for map 3 is obtained.

\begin{figure}[t!]
	\centering
%
%
\definecolor{mycolor1}{rgb}{0.00000,0.44700,0.74100}%
\definecolor{mycolor2}{rgb}{0.85000,0.32500,0.09800}%
\definecolor{mycolor3}{rgb}{0.92900,0.69400,0.12500}%
\begin{tikzpicture}

\begin{axis}[%
legend style={font=\fontsize{6}{5}\selectfont},
width=8cm,
height=6cm,
bar shift auto,
symbolic x coords={11,25,33},
enlarge x limits=0.2,
xtick=data,
xlabel={Selected map's building density [\%]},
ymin=0,
ymax=30,
ylabel={AP selection error [\%]},
xmajorgrids,
ymajorgrids,
legend style={at={(0.03,0.97)}, anchor=north west, legend cell align=left, align=left, draw=white!15!black}
]
\addplot[ybar, fill=mycolor1, draw=black, area legend] table[row sep=crcr] {%
11	5.7\\
25	10.9\\
33	27\\
};
\addplot[forget plot, color=white!15!black] table[row sep=crcr] {%
11	0\\
33	0\\
};
\addlegendentry{Switching 1 AP}

\addplot[ybar, fill=mycolor2, draw=black, area legend] table[row sep=crcr] {%
11	5.4\\
25	10\\
33	23.1\\
};
\addplot[forget plot, color=white!15!black] table[row sep=crcr] {%
11	0\\
33	0\\
};
\addlegendentry{Switching 2 APs}

\addplot[ybar, fill=mycolor3, draw=black, area legend] table[row sep=crcr] {%
11	1.45\\
25	3.4\\
33	13.3\\
};
\addplot[forget plot, color=white!15!black] table[row sep=crcr] {%
11	0\\
33	0\\
};
\addlegendentry{Switching 3 APs}

\end{axis}
\end{tikzpicture}%
	\caption{AP selection error based on the predicted LSF of the off APs, calculated over all possible outdoor locations of the UE demanding extra SE, assuming full REMs of active APs are available.}
	\label{1mapAPsel}
\end{figure}
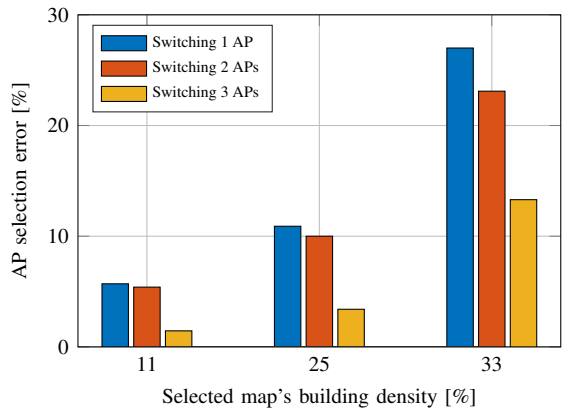

\subsubsection{AP Selection Performance}
The AP selection error is evaluated in Figure \ref{1mapAPsel} for a scenario in which the full REM map of 70 active APs is available. In this case, we perform AP selection of the best one, two, and three APs from 10 sleep APs across all outdoor map locations. As shown in the figure, a maximum AP selection error of approximately 10\% is achieved in map 2 and map 3. This means that for 90\% of user locations, we are able to switch the optimal set of APs without a need to switch any of the off APs to do channel gain measurements. In the case of map 1, which has a higher building density compared to map 2 and map 3, an AP selection error of 27\% is obtained when only one extra AP is needed, whereas in case 3 extra APs are needed to meet the user's SE, the AP selection error goes down to 12\%. This means that for the case where a single AP is needed at 15\% of the locations, the second-best or third-best AP is turned on.  

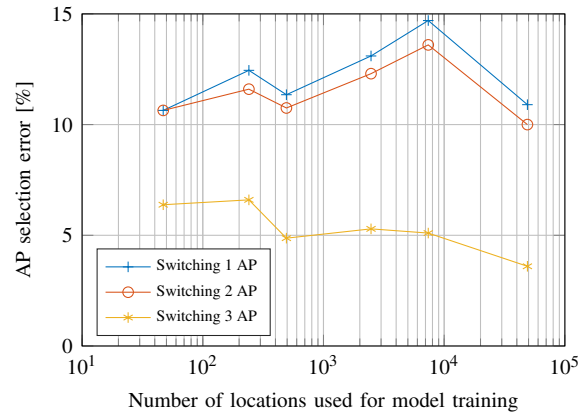
\begin{figure}[t!]
	\centering
%
%
\definecolor{mycolor1}{rgb}{0.00000,0.44700,0.74100}%
\definecolor{mycolor2}{rgb}{0.85000,0.32500,0.09800}%
\definecolor{mycolor3}{rgb}{0.92900,0.69400,0.12500}%
\begin{tikzpicture}

\begin{axis}[%
legend style={font=\fontsize{6}{5}\selectfont},
width=8cm,
height=6cm,
xmode=log,
xmin=10,
xmax=100000,
xminorticks=true,
xlabel={Number of locations used for model training},
ymin=0,
ymax=15,
ylabel={AP selection error [\%]},
xmajorgrids,
xminorgrids,
ymajorgrids,
legend style={at={(0.03,0.03)}, anchor=south west, legend cell align=left, align=left, draw=white!15!black}
]
\addplot [color=mycolor1, mark=+, mark options={solid, mycolor1}]
  table[row sep=crcr]{%
48942	10.9\\
7381	14.7\\
2477	13.1\\
493	11.35\\
241	12.45\\
47	10.64\\
};
\addlegendentry{Switching 1 AP}

\addplot [color=mycolor2, mark=o, mark options={solid, mycolor2}]
  table[row sep=crcr]{%
48942	10\\
7381	13.6\\
2477	12.3\\
493	10.75\\
241	11.6\\
47	10.64\\
};
\addlegendentry{Switching 2 AP}

\addplot [color=mycolor3, mark=asterisk, mark options={solid, mycolor3}]
  table[row sep=crcr]{%
48942	3.6\\
7381	5.1\\
2477	5.29\\
493	4.87\\
241	6.6\\
47	6.38\\
};
\addlegendentry{Switching 3 AP}

\end{axis}
\end{tikzpicture}%
	\caption{AP selection error based on the predicted LSF of the off APs versus the number of locations/pixels used to train our u-net model for a map with a building density of 25\%.}
	\label{APselLimitUsr}
\end{figure}

Since the assumption of having the full REM of active APs might be strict, in Figure \ref{APselLimitUsr}, we present a case where only the \ac{LSF} measurements at selected locations are available in map 2. Such measurements can be collected throughout the time window at which the active APs are on. We use the same locations to predict the \ac{LSF} of sleep APs. These locations are randomly chosen to represent 15\%, 5\%, 1\%, 0.5\%, and 0.1\% of the total number of pixels (256 $\times$ 256), which corresponds to the number of outdoor locations in map 2 of 7381, 2477, 493, 241, and 47, respectively. These different percentages of outdoor locations are chosen to asses our REM model with a wide range of outdoor locations, i.e., from tens to thousands of locations. The figure also presents the case where all outdoor locations in map 2 are considered, which corresponds to 48942 outdoor locations. As shown in the figure, an AP selection error lower than 15\% is guaranteed for cases where one, two, or three extra APs are needed, regardless of the considered number of outdoor locations.

\section{Conclusion} \label{sec:conclusion}

We presented an accurate time-efficient \ac{REM} prediction framework based on u-net \ac{CNN}. We investigated several data preprocessing steps and quantified their impact on the \ac{RMSE} of predicted REMs. The presented preprocessing steps include three different approaches for fractional LoS maps calculation, building-density-based data split, data augmentation, as well as the u-net model loss function. We evaluated the performance of our proposed framework using the 3D city maps from the \textit{RadioMap3DSeer} dataset and highlighted the performance gain and corresponding tradeoffs. In particular, the performance of our proposed framework has been evaluated in the context of \textit{the 2023 IEEE ICASSP Signal Processing Grand Challenge, namely, the First Pathloss Radio Map Prediction Challenge}. Our results have shown that the proposed framework provides a normalized average \ac{RMSE} of 0.045 on the challenge's test set, with an average runtime of 14 milliseconds per map. Finally, a relevant CF-mMIMO use case is presented, in which we demonstrated that one could obviate consuming energy on large-scale fading measurements and rely on predicted REM instead to select which APs to switch on. In particular, we showed that by exploiting predicted \ac{REM} an \ac{AP} selection error of around 5\% in case a \ac{UE}'s \ac{SE} needs three extra \acp{AP} is achieved.

\bibliographystyle{IEEEbib}
\balance
\bibliography{refs}

\vfill\pagebreak

\end{document}